\documentclass[a4paper,12pt]{article}
\usepackage{comment}
\usepackage{jheppub}
\usepackage{dsfont}
\usepackage{amsmath,amssymb,amscd,amsfonts,mathtools}
\usepackage{amsthm,thmtools}
\usepackage{xcolor}
\usepackage{tikz}
\usepackage{tikz}
\usepackage{amsmath}
\usepackage{comment}
\usepackage{empheq}
\usepackage{mathrsfs} % for \mathscr
\usetikzlibrary{decorations.pathmorphing}

\preprint{UT-WI-33-2025}
\title{On  thermal holographic RG flows}

\author[]{Elena C\'aceres and }
\author[]{Hare Krishna}

\affiliation[]{Theory Group, Weinberg Institute, Department of Physics, University of Texas at Austin,\\
2515 Speedway, Austin, Texas 78712, USA.}

\emailAdd{elenac@utexas.edu}
\emailAdd{hkrishna.phy@gmail.com}

\abstract{ Holographic Renormalization Group (RG) flows, described by Einstein gravity coupled to matter fields, have been thoroughly explored in the context of vacuum states. In this work, we shift the focus to \emph{thermal} states. We develop the Hamilton-Jacobi functional up to \emph{second order} in the boundary derivative expansion. Using this formalism, we derive the Ward identity associated with broken dilatation symmetry for thermal correlators. Furthermore, we also obtain the modified Callan-Symanzik equation for the dual thermal CFT. 
We then employ the Hamiltonian constraint approach to analyze the black hole interior and comment on the near-singularity behavior from this perspective.  %This approach provides a relation between dynamics near the singularity and boundary coupling.
}

\begin{document}
\maketitle
    \section{Introduction  }
The study of renormalization group (RG) flows in conformal field theories (CFTs) is a well-established and extensively developed subject. A CFT, specified by its set of local operators with scaling dimensions $\Delta_O$ and structure constants $f_{ijk}$, can be deformed by a relevant operator of dimension $\Delta_O < d$. Such a deformation triggers a RG flow in which the operator dimensions and structure constants evolve, driving the theory from a UV fixed point to an IR fixed point. At the IR endpoint, the CFT is characterized by a different effective set of operators and couplings. The central charge, which counts the effective degrees of freedom, also decreases along the flow. This monotonic behavior was first established in two dimensions by Zamolodchikov \cite{Zamolodchikov:1986gt}, extended to four dimensions by Komargodski et al. \cite{Komargodski:2011vj,Cardy:1988cwa}, and later generalized to other dimensions \cite{Klebanov:2011gs,Jafferis:2011zi}. Entropic proofs of these $c$-theorems have been provided by various authors as well \cite{Casini:2004bw,Casini:2012ei}.  

AdS/CFT offers a different perspective: RG flows correspond to bulk gravitational dynamics. In this framework, monotonic functions along the flow can be derived holographically by demanding that the background satisfies the Null Energy Condition  (NEC) \cite{Myers:2010tj,Myers:2010xs}.  A particularly intriguing feature arises when considering \emph{thermal} states, dual to black holes in the bulk. In this case, from the bulk point of view, it is natural to continue the flow into the black hole interior \cite{Frenkel:2020ysx, Caceres:2021fuw, Caceres:2022smh}. A  holographic $a$-function can be obtained, and it has been shown to continue to decrease in the interior \cite{Caceres:2022smh, Caceres:2022hei, Caceres:2023zft, Caceres:2024edr} but the precise boundary interpretation of this flow remains an open question. The present work, however, focuses on a different facet of thermal RG flows.

Consider a thermal CFT defined on the boundary manifold $ M=S^1_{\beta} \times R^{d-1}$. 
    At finite temperature, the conformal symmetry of the boundary CFT is broken. %down to its subgroup. 
     The remaining symmetries are translation and rotation along the spatial direction. %are left unbroken. 
     The breaking of the symmetry is compensated by a variation with another scale, namely, the temperature $T=1/\beta$. Thus, the Ward identity associated with broken dilatation is 
     \begin{align}
    \label{boxedeq1}
    \boxed{
        (D+\beta \frac{\partial}{\partial \beta})\langle O_1(x_1)\cdots O_n(x_n)\rangle_{\beta}=0
        }
    \end{align}
    Here, $D$ stands for the dilatation operator. We can write similar Ward identities for other broken symmetries  \cite{Marchetto:2023fcw}. Furthermore, one can also add a relevant deformation. Intuitively, the above Ward identity will contain additional terms involving the beta function for the coupling as $\beta_{\phi_0} \frac{\partial}{\partial \phi_0}$. The complete flow in temperature and coupling space is yet to be understood. But an infinitesimal deformation by a relevant operator is tractable. 
    
    As is standard in AdS/CFT,  the thermal state is dual to a black hole, and the deformation by a relevant operator is dual to a scalar field in the asymptotically AdS spacetime.  The relevant deformation drives the system to the IR, and the radial direction provides a measure of the energy scale of the boundary theory. Thus, the UV fixed point corresponds to the $r\rightarrow \infty$ limit of the radial direction (the boundary), while the IR corresponds to something deep in the bulk. One can understand many aspects of CFTs such as anomalies, RG flows, etc \cite{Skenderis:1999mm,Skenderis:2002wp} using holographic methods. Many results have been obtained assuming the full conformal symmetries of CFT, i.e., at zero temperature. But finite temperature backgrounds, where some of the symmetries are broken, are less explored. Our work aims to fill in some of these gaps \footnote{There are some earlier works on RG flows in SYK-like models at finite temperature by Anninos et.al.\cite{Anninos:2020cwo,Anninos:2022qgy}. Aspects of thermalization and hydrodynamics are discussed in \cite{Delacretaz:2021ufg}. A constraint on strongly coupled gauge theories at finite temperature is discussed in \cite{Appelquist:1999hr,Zabzine:1997gh}. }.

One of our main results is to derive, holographically, the ``modified'' Callan-Symanzik equation for the thermal boundary theory, generalizing the seminal work of de Boer, Verlinde, and Verlinde\cite{deBoer:1999tgo}. To do so,  we study the dilatation Ward identity \eqref{boxedeq1} from the holographic perspective. We show that the Hamiltonian constraints of the bulk yield the thermal  Callan-Symanzik equation for boundary theory, while the momentum constraint establishes the translational invariance of the boundary theory. We extend their analysis to a systematic derivative expansion at fixed temperature parameter $D = T s$, where $T$ and $s$ are the temperature and entropy density of the black brane. The resulting local Hamilton characteristic functional is determined up to two derivatives in boundary coordinate. This development is presented in section~\ref{sec:thermal-hj-derivative-expansion}.

Another contribution of this work is related to the black hole interior and is presented in section~\ref{interior}. We study the dynamics near the singularity and derive the Hamiltonian and momentum constraints for the interior. It is known that near the singularity, the metric and scalar fields exhibit, generically, rich dynamics known as Mixmaster dynamics \cite{LIFSHITZ1992659,Belinsky:1970ew,PhysRevLett.22.1071} and are governed by the Belinski–Khalatnikov–Lifshitz (BKL) exponents. Using Hamilton-Jacobi theory, we show that the constraints we derive give rise to a functional equation for the on-shell action in terms of BKL exponents. We comment on the resemblance of these equations to those encountered in quantum cosmology.  \cite{PhysRev.186.1319}. 

This article is organized as follows.  In section \ref{superpotential section}, we review some of the aspects of gravitational and scalar field equations. In particular, the superpotential formalism allows Einstein's equations to be written as a system of first-order equations. In section \ref{sec:thermal-hj-derivative-expansion}, we formulate the Hamilton-Jacobi theory at finite temperature and find Hamiltonian and momentum constraints. We later recast these constraints as the Callan-Symanzik equation for the boundary correlation function at finite temperature. Next, in section \ref{interior}, we investigate the interior of the black hole using the Hamilton-Jacobi equations. This gives rise to a functional equation for the on-shell action in terms of the BKL exponents near the singularity.
Our work opens up several lines of research that we discuss in section \ref{discussions}. Appendix \ref{app:thermal-HJ} contains the formalism of Hamilton-Jacobi theory at any temperature, and  Appendix \ref{ward identity} is a discussion of the Ward identity for broken conformal symmetry due to the finite temperature. \\

\textbf{Note on Notation:}

\begin{itemize}
    \item Greek indices \(\mu, \nu, \ldots\) denote hypersurface coordinates- time, and spatial directions, i.e., \(\mu = \{t, x, y\}\).
    \item Latin indices from the middle of the alphabet \(i, j, \ldots\) denote spatial coordinates on the hypersurface, i.e., \(i = \{x, y\}\).
    \item Latin indices from the beginning of the alphabet \(a, b, \ldots\) are used for bulk spacetime coordinates, which generally include the hypersurface directions ($\mu=\{t,x,y\}$) and a radial direction $z$ or $u$.
\end{itemize}

    \section{Preliminaries}
    \label{superpotential section}
   % In this section, we study the bulk dual of RG flows in the thermal state.  The boundary theory lies on the $S^1 \times R^{d-1}$ manifold. To understand the RG flow, we perturb the theory with a relevant operator $(\Delta<d=3)$, and we have a non-trivial RG flow from UV theory (original CFT) to IR theory. The bulk dual action includes Einstein-Hilbert and the scalar fields' contributions. We seek a black hole with a scalar field solution to the coupled system. Here we are using the AdS/CFT dictionary of scalar operators are dual to the scalar fields in the bulk, and the radial coordinate serves as the energy scale.  \\
    
%The RG flows in the vacuum state have been widely explored. A lot of results like holographic C theorems, entanglement entropy, are understood both from a holographic and purely from CFT point of view  \cite{Ryu:2006bv,Ryu:2006ef,Calabrese:2004eu,Vidal:2002rm,Komargodski:2011vj,Casini:2004bw,Casini:2011kv,Casini:2012ei,Batrachenko:2004fd,Papadimitriou:2004ap,Papadimitriou:2005ii,Skenderis:2002wp,Skenderis:1999mm,Erdmenger:2001ja,Kalkkinen:2001vg,Balasubramanian:2001nb,Kiritsis:2014kua}. But the 
Holographic RG flow in thermal states has not been as thoroughly explored in the literature as its vacuum counterpart. One of the first developments \cite{Gursoy:2018umf} studied the effect of different potentials for the scalar field. Their work uses the ``superpotential'' formalism that we will review shortly. Another line of work emphasized extending the RG flow beyond the horizon to investigate the interior of black holes.  However, neither of these approaches explored the Hamilton-Jacobi (HJ) analysis. In this section, we will first review the work of G\"ursoy et al. \cite{Gursoy:2018umf} and find the superpotential equations describing the flow. When we turn to the HJ analysis in the next section, we will show that HJ reproduces the superpotential equations. 

    \subsection{Setup}
Consider a boundary theory that lies on $S^1 \times R^{d-1}$ manifold. To understand the RG flow, we perturb the theory with a relevant operator $(\Delta<d=3)$, and this perturbation triggers an RG flow from UV theory (original CFT) to IR theory. The bulk dual of this flow involves an Einstein-Hilbert plus scalar action. To find a solution compatible with the symmetries of the problem and working in Euclidean signature,  we take the ansatz,%\footnote{
%Note that the anisotropic scaling in the $t$ and $x$ directions reflects the fact that in the CFT we have a distinguished direction $t$ with periodicity $\beta$, which breaks the conformal symmetry to its subgroup.}

     \begin{eqnarray}
\label{metphiansatz1}
    g_{\mu\nu} dx^{\mu}dx^{\nu}= \frac{1}{f(z)} dz^2+ e^{2 A(z)}(f(z) dt^2+ dx^i dx^i), \quad \phi= \phi(z)
\end{eqnarray}
Here $f(z)$ is the usual blackening factor for the black hole. Note that the anisotropic scaling in the $t$ and $x$ directions reflects the fact that in the CFT we have a distinguished direction $t$ with periodicity $\beta$, which breaks the conformal symmetry to its subgroup. Note that in the Lorentzian signature, we have an interior ($f(z)<0$) and exterior with $f(z)>0$. The zeros of $f(z)$ define the horizon. The Radial Hamiltonian evolves the time-like slice in the interior (from horizon $z=z_h$ to singularity $ z=\infty$) and evolves the spacelike slice at the outside (from $z<z_h$ to boundary $z=-\infty$). One can consider both sets of evolution and glue them at the horizon.  It is often more convenient to use the infalling coordinates, which are smooth near the horizon. The metric in in-falling coordinates can be written as
\begin{eqnarray}
\label{metinfalling}
    ds^2= -e^{2 A(z)} f(z) du^2+ e^{A(z)} 2 du dz +e^{2 A(z)} dx^i dx^i, \quad \phi= \phi(z) 
\end{eqnarray}
One can scale the radial coordinate $z$ to absorb the factor of $e^{A(z)}$ in the metric component $g_{uz}$. 
Lorentzian signature will become relevant when we study the black hole interior in section \eqref{interior}, but for the moment, let us focus on  Euclidean signature. 
%It is interesting to develop the Hamilton-Jacobi theory directly in the infalling coordinate.}.

We consider two derivative action of gravity, which are coupled to matter fields (in this case, it is a scalar) with an arbitrary potential $V(\phi)$. In the above equation, we consider the scalar field dependence only on the radial coordinate $z$.

\begin{eqnarray}\label{eq:Euclidean_action}
    S_{gr}=-\frac{1}{16 \pi G} \int d^4 x \sqrt{g}\Big[ R-\frac{1}{2} g^{ab} \partial_{a}\phi \,\partial_{b}\phi- V(\phi)\Big]+S_{GHY}+S_{CT}
\end{eqnarray}
The boundary term can be written as
\begin{eqnarray}
    S_{GHY}=-\frac{1}{8 \pi G} \int d^3 x \sqrt{-\gamma}\, K,
\end{eqnarray}
where the signs of various terms are consistent with the Euclidean signature\footnote{In Lorentzian signature, the matter action can be written as
\begin{eqnarray}
 S_{matter}=\int \sqrt{-g}\, d^4 x\,\ \Big(-\frac{1}{2} g^{ab} \partial_{a}\phi \,\partial_{b}\phi- V(\phi)\Big)   \end{eqnarray}}.
 The potential also contains the cosmological constant term along with the mass term and higher order terms as $V(\phi)\sim \frac{-6}{\ell^2}+\frac{1}{2} m^2 \phi^2 +\cdots $. And $S_{CT}$ denotes the usual  counterterms needed for  holographic renormalization \cite{Skenderis:1999mm,Balasubramanian:1999re}.\\

It is straightforward to obtain  Einstein equations,  $ R_{ab}- \frac{1}{2} g_{ab} R= 8 \pi\, G\, T_{ab}^{\, matter}$, fo the action \eqref{eq:Euclidean_action} with the matter stress tensor given by,
\begin{eqnarray}
T_{ab}^{\, matter}= \nabla_{a} \phi \, \nabla_{b} \phi- \frac{1}{2} g_{ab} g^{cd} \nabla_{c}\phi\,\nabla_{d}\phi- g_{ab}\, V(\phi).
\end{eqnarray}

For the metric and scalar ansatz \eqref{metphiansatz1}, we obtain, 
\begin{eqnarray}
  &&E_{zz}=\frac{A'(z) f'(z)+f(z) \left(3 A'(z)^2-4 \pi 
   G \phi '(z)^2\right)+8 \pi  G V}{f(z)} =0\\
   &&E_{tt}=-e^{2 A(z)} f(z) \left[A'(z) f'(z)+f(z)
   \left(2 A''(z)+3 A'(z)^2+4 \pi  G \phi
   '(z)^2\right)+8 \pi  G V\right]=0\nonumber\\
   &&E_{xx}=\frac{1}{2} e^{2 A(z)} \big[5 A'(z)
   f'(z)+f(z) \left(4 A''(z)+6 A'(z)^2+8 \pi 
   G \phi '(z)^2\right)+f''(z)+\nonumber\\
   &&\hspace  {4.8in}16 \pi  G V\big]=0\nonumber\\
   &&E_{xx}=E_{yy}=0\nonumber
\end{eqnarray}
Rewriting these equations in a more convenient form and setting  $\frac{1}{16 \pi G}\equiv1$, we have,  %these equations as\footnote{These manipulations are not essential, but we get a much simpler equation to deal with, and these can be written in terms of superpotential. We have set $\frac{1}{16 \pi G}=1$.},
\begin{eqnarray}
&&  E_{tt}  -\frac{1}{2} E_{xx}=\frac{1}{2} \left(-3 A'(z) f'(z)-f''(z)\right)=0\label{eq:ett_minus_exx}\\
&&E_{xx}-E_{zz}= \left(4 A''(z)+\phi '(z)^2\right)=0 \label{eq:Wsolves}\\
&&E_{zz}=2 f'(z) A'(z) +\Big(6 A^{'2}(z)-\frac{1}{2} \phi^{'2}(z)\Big)f+V(\phi)=0.
\end{eqnarray}
Finally, the equation of motion for the scalar field is,
\begin{eqnarray}
 \frac{1}{\sqrt{-g}}\partial_{a}(\sqrt{-g} g^{ab} \partial_{b} \phi)=   f(z) \left[3 A'(z) \phi '(z)+\phi ''(z)\right]+f'(z)
   \phi '(z)-\frac{\partial V(\phi)}{\partial \phi}&&=0.\nonumber\\
   && 
\end{eqnarray}

% The equation of motion for the scalar field can be written as
% \begin{eqnarray}
%  \frac{1}{\sqrt{-g}}\partial_{\mu}(\sqrt{-g} g^{\mu\nu} \partial_{\nu} \phi)=   f(z) \left(3 A'(z) \phi '(z)+\phi ''(z)\right)+f'(z)
%    \phi '(z)-\frac{\partial V(\phi)}{\partial \phi}=0
% \end{eqnarray}
The metric ansatz \eqref{metphiansatz1} describes a black hole having a horizon at $f(z_h)=0$, which has temperature and entropy density as
\begin{eqnarray}
    T=\frac{e^{A(z_h)}}{4 \pi}|f'(z_h)|, \quad s= 4 \pi M_p^2 e^{2 A(z_h)}.
\end{eqnarray}

The AdS Schwarzchild black hole is a UV fixed point. Here we take $\phi=\phi_0$ and $V'(\phi)=0$. The AdS length scale is $\ell=\sqrt{-6/V(\phi_0)}$. Then the black hole solution can be written with its corresponding dependencies as
 \begin{eqnarray}
     A(z)= -\frac{z}{\ell}, \quad f(z)= 1- e^{3 (z-z_h)}/\ell, \quad \phi(z)= \phi_0.
 \end{eqnarray}
The temperature and entropy density for this AdS black hole solution are
 \begin{eqnarray}
     T= \frac{3}{4 \pi \ell} e^{- z_h/\ell}, \quad s=4 \pi M_p^{2} e^{-2z_h/\ell}.
 \end{eqnarray}
 This also defines the temperature and entropy density at the UV fixed point.\\
 
 Let us integrate the $E_{tt}  -\frac{1}{2} E_{xx}$ Einstein equation \eqref{eq:ett_minus_exx}. We get, 
\begin{eqnarray}
    f'(z)e^{3 A(z)}=-D
    \label{eq:D definition}
\end{eqnarray}
Here we are doing the integration from the boundary to the horizon. % Here, it is just showing that the derivatives of $f(z)$ and $e^{3 A(z)}$ are related to each other. 
The integration constant $D$ can be related to the entropy and temperature at the horizon $z=z_h$ as
\begin{eqnarray}
     \frac{T \,s}{M_p^2}= D
     \label{D-definition}
\end{eqnarray}
In the UV theory, we have asymptotic AdS with a potential as
\begin{eqnarray}
    V(\phi) \sim -\frac{6}{\ell^2}+\frac{m^2\phi^2} {2}
\end{eqnarray}
In the asymptotic AdS region ($z \rightarrow -\infty$), the solution has the familiar fall-off as
\begin{eqnarray}
  &&  \phi= \phi_- e^{\Delta_- z/\ell}+\phi_+ e^{\Delta_+ z/\ell}+\cdots, \quad \Delta_{\pm}=\frac{3}{2}\pm \frac{1}{2} \sqrt{9+4 m^2 \ell^2}\\
  &&  A(z)=-\frac{z}{\ell}+\cdots, \quad f(z)= 1- \frac{\ell D}{3} e^{3 z/\ell}+\cdots
\end{eqnarray}
The parameters in the scalar field fall off are related to the coupling $\phi_0$ (UV coupling) and the vev of scalar operators as
\begin{eqnarray}
    \phi_-= \phi_0\,  \ell^{\Delta_-}, \quad \phi_+=\frac{\langle O \rangle \ell^{\Delta_+}}{(3- 2\Delta_-)(M_p \ell)^2}
\end{eqnarray}
In standard quantization, the CFT flows are driven by deforming them with a relevant operator. %In this subsection, we set up our convention and write the Einstein equation. In the next subsection, we write the Einstein equation in terms of a superpotential.

\subsection{First order formalism}
The first order formalism to solve  Einstein's equations relies on defining an auxiliary function, called the superpotential, $W$.
This method reduces the problem of solving the full second-order equations of motion to solving a set of simpler, first-order equations\footnote{At zero temperature, these are just the BPS equations.}.

More precisely, if we can define $W(\phi)$, such that,   
\begin{eqnarray}
\label{s1met}
 A'[z]= -\frac{1}{4} W[\phi(z)], \quad \phi'(z)=\frac{d W[\phi(z)]}{d \phi}
\end{eqnarray}
then we have solved equation \eqref{eq:Wsolves}. %%\footnote{At zero temperature, one can write down the action in terms of a sum of squares, and it resembles the BPS equation. This is why we are calling $W(\phi)$ as superpotential.}.
\begin{eqnarray}
    A''(z)+\frac
   {1}{4}\phi'(z)^2= -\frac{1}{4}\frac{d W[\phi(u)]}{d \phi} \phi'+\frac{1}{4}(\frac{d W[\phi]}{d \phi})^2=0.
\end{eqnarray}
Here, the relationship between the superpotential $W(\phi)$ and metric coefficients \eqref{s1met} is the same as in the zero temperature case. We can also write the scale factor as a functional of $\phi$ as
\begin{eqnarray}
    \frac{d A}{d \phi}= \frac{d A}{d z} \frac{ dz}{ d \phi}= -\frac{1}{4} \frac{W(\phi)}{\partial_{\phi} W(\phi)}
\end{eqnarray}
The radial dependence of $\phi(z)$ can be thought of as the RG flow of the boundary theory coupling. And the scale factor $e^{A(z)}$ can be thought of as an energy scale $\mu$. 
\begin{eqnarray}
    \mathrm{log} \, \mu \equiv A(z)
\end{eqnarray}
Then the beta function of the theory can be written as
\begin{eqnarray}
   \beta(\phi)\equiv  \frac{ d \, \phi}{ d \, \mathrm{log}\, \mu}= -4 \frac{\partial_{\phi} W(\phi)}{W(\phi)}.
\end{eqnarray}
Note that  $W[\phi(z)],\, f[\phi(z)]$ is  a functional of $\phi(z)$ which depends on the radial coordinates $z$. We can plug this into other Einstein equations, and we get
\begin{eqnarray}
\label{eq1}
\left(3 A'(z) f'(z)+f''(z)\right)=\frac
 {d^2 f}{d \phi^2}\frac{d W}{d \phi}+\frac{d f}{d \phi} \frac{d^2 W}{d \phi^2}-\frac{3}{4} W \frac{d f}{d \phi}=0
\end{eqnarray}
\begin{eqnarray}
\label{superpotential1check}
\Bigg(\frac{1}{2}\Big(\frac{d W}{d \phi}\Big)^2-\frac{3}{8} W^2\Bigg)f+\frac{W}{2} \frac{d f}{d \phi} \frac{d W}{d \phi}-V(\phi)=0
\end{eqnarray}

The above equation relates the superpotential to the potential of the scalar fields coupled to gravity. In the zero temperature case $f(\phi)=1$, it gives the familiar relationship as 
\begin{eqnarray}
  \Bigg(\frac{1}{2}\Big(\frac{d W}{d \phi}\Big)^2-\frac{3}{8} W^2\Bigg)= V(\phi). 
  \label{superpotentialrelation}
\end{eqnarray}
\textbf{Boundary conditions}\hfill\break
The system of equations \eqref{eq1} and \eqref{superpotential1check} is effectively a third-order equation. And we need three integration constants to completely specify the solution. These constants could be understood as the boundary value problem for the $f(\phi)$ and the initial value problem for the super-potential $W(\phi)$. This can fix the integration constants as
\begin{eqnarray}
    f(\phi=0)=1, \quad f(\phi=\phi_h)=0, \quad W(\phi=\phi_h)= W_h
\end{eqnarray}
The first equation is the statement about the UV behavior of field theory, where we have a conformal fixed point. The second equation specifies the horizon of the solution. The $W_h$ can be written in terms of horizon position in the RG flowed theory parametrized by $\phi_h$. Hence, we have effectively a one-parameter family of solutions. With these boundary conditions, we solve these equations for the superpotential.\\

\noindent\textbf{Outside the horizon}\hfill\break
The \eqref{eq1} can also be written as
\begin{eqnarray}
\label{eq1sim}
    \frac{d}{d \phi}\Big[\frac{d f}{d \phi}\frac{d W}{d \phi}e^{3 A_1(\phi)}\Big]=0, \quad \text{where}\,\, A_1(\phi)\equiv -\frac{1}{4} \int^{\phi} d\hat{\phi} \frac{W}{W'}
\end{eqnarray}
We can integrate the above equation, and we get
\begin{eqnarray}
\label{eq1simsol1}
    \frac{d f}{d \phi}= - \frac{\tilde{D}}{W'} e^{- 3 A_1(\phi)}
\end{eqnarray}
The additive constant ambiguity in $A_1(\phi)$ can be fixed by requiring that $A_1(\phi_h)=0$, And then we have
\begin{eqnarray}
   A_1(\phi)= -\frac{1}{4} \int^{\phi}_{\phi_h} d\hat{\phi} \frac{W}{W'}
\end{eqnarray}
The simplified equation \eqref{eq1simsol1} can further be written as
\begin{eqnarray}
    f(\phi)= \tilde{F}-\tilde{D}\int_{0}^{\phi}\frac{dy}{W'(y)} e^{-3 A_1(y)}
\end{eqnarray}
In the UV where $\phi \rightarrow 0$, we have $f(\phi)=1$ and that fixes the constant $\tilde{F}=1$ and other constant
\begin{eqnarray}
    \tilde{D}= \Bigg(\int_0^{\phi_h} \frac{dy}{W'(y)} e^{-3 A_1(y)}\Bigg)^{-1}
\end{eqnarray}
Now we can evaluate the equation \eqref{superpotential1check} at $\phi=\phi_h$ and we get
\begin{eqnarray}
    W(\phi_h)\equiv W_h= -2 \frac{V(\phi_h)}{\tilde{D}} 
\end{eqnarray}
Here we have established that the superpotential $W(\phi_h)$ depends on the ordinary potential. This justifies the one-parameter family of solutions. In addition, the superpotential satisfies the following properties:
\begin{enumerate}
    \item If we assume the regularity at the horizon, then the functions $W(\phi)$ and $f(\phi)$ are completely specified by a single parameter $\phi_h$.
    \item The superpotential is a monotonically increasing function along the flow as a function of the radial coordinate $z$. But it can be multivalued as a function of $\phi$.
    \begin{eqnarray}
        \frac{d W(z)}{d z}=\frac{d \phi(z)}{d z} \frac{d W(\phi)}{d \phi}= W^{'2} \geq 0
    \end{eqnarray}
    \item The fact that  $W(z)$ is a monotonically increasing  function of $z$ is related to a holographic  c-theorem. Indeed, using the null energy condition, $T_{\mu\nu}k^{\mu}k^{\nu}\geq 0,$  yields\footnote{We choose $k_{\mu}=\left\{1,-\frac{1}{2}e^{A(z)} f(z),0,0\right\}$}, 
    \begin{eqnarray}
  -\frac{1}{2} f(z)^2 A''(z) \geq 0 \quad \implies A''(z)\leq 0
  \end{eqnarray}
  which is consistent with the monotonicity of $W(\phi)$ since,  $A'[z]= -\frac{1}{4} W[\phi(z)]$ and thus,  $A''[z]= -\frac{1}{4} \frac{d W}{d \phi} \frac{d \phi}{d z} \leq 0$
  %\subsection*{Null energy constraints and monotonicity}

   % One can find a monotonic function using the null energy condition starting from the work of Myers et al. \cite{Myers:2010tj,Myers:2010xs}. The idea is straightforward to implement. Using Einstein's equation, one finds the stress tensor corresponding to the background. And then one uses the null energy condition.
%$$T_{\mu\nu}k^{\mu}k^{\nu}\geq 0$$
%For the metric ansatz \eqref{metphiansatz1}, It gives
%\begin{eqnarray}
 % -\frac{1}{2} f(z)^2 A''(z) \geq 0 \quad \implies A''(z)\leq 0
%  \end{eqnarray}
%Hence, the monotonic decreasing function would be $A'(z)$. The null vector that we chose is
%\begin{eqnarray}
%k_{\mu}=\left\{1,-\frac{1}{2}e^{A(z)} f(z),0,0\right\}
%\end{eqnarray}
%This is also consistent with above C theorem by remembering $A'[z]= -\frac{1}{4} W[\phi(z)]$. The second derivative $A''[z]= -\frac{1}{4} \frac{d W}{d \phi} \frac{d \phi}{d z} \leq 0$
\end{enumerate}
% \subsection*{Lower bounds on superpotential}
% We can algebraically solve the \eqref{superpotential1} equation for superpotential $W(\phi)$. By Null energy constraints and the holographic C theorem, we can assume the positivity of the superpotential and $A'(z)\leq 0$. Now it gives
% \begin{eqnarray}
%     W(\phi)=\frac{2}{3} \frac{f'W'}{f}+\sqrt{(\frac{2}{3} \frac{f' W'}{f})^2+\frac{4}{3}(W'^2-\frac{2 V}{f})}
% \end{eqnarray}
% As long as the potential is negative, there is always a real solution for $W(\phi)$ at the critical points i.e. $W'=0$. It also implies that the superpotential $W(\phi)>0$ as long as the potential is negative. In general, for the real solution for $W$, the existence condition can be written as
% \begin{eqnarray}
%     \Big(\frac{f' W'}{f}\Big)^2 \geq 3\Big(W'^2-\frac{2 V}{f}\Big)
% \end{eqnarray}
% At the critical points where $W'=0, \frac{f' W'}{f}=0$, then we have
% \begin{eqnarray}
%     W=\sqrt{-\frac{8}{3} \frac{V}{f}}
% \end{eqnarray}
% This is the same as the zero temperature case, where we have set $f=1$.\\

% \textbf{At the horizon:-} To understand the behavior of superpotential at the horizon, we need to set $\phi=\phi_h$ and $f(\phi_h)=0$ then the superpotential eq \eqref{superpotential} becomes
% \begin{eqnarray}
%    W(\phi_h)= 2 \frac{V(\phi_h)}{df/d\phi\,\, dW/d\phi}\Big|_{\phi=\phi_h} 
% \end{eqnarray}

\begin{figure}
\centering
\begin{tikzpicture}[scale=2.7]
\draw[-,draw=none,fill=black!10] (1,1) to (1,-1) to (0,0) to (1,1); 
\draw[-,draw=none,fill=black!10] (-1,1) to (-1,-1) to (0,0) to (-1,1);
\node at (1.1,0) {R};
\node at (-1.1,0) {L};
\draw[-,draw=none,fill=red!10] (-1,1) ..controls (-0.3,0.8) and (0.3,0.8) .. (1,1) to (0,0) to (-1,1); 
\draw[-,draw=none,fill=red!10] (1,-1) ..controls (0.3,-0.8) and (-0.3,-0.8) .. (-1,-1) to (0,0) to (1,-1);

\draw[-,thick,red,decoration = {zigzag,segment length = 1mm, amplitude = 0.25mm},decorate] (-1,1) ..controls (-0.3,0.8) and (0.3,0.8) .. (1,1);
\draw[-,thick] (1,1) to (1,-1);
\draw[-,thick,red,decoration = {zigzag,segment length = 1mm, amplitude = 0.25mm},decorate] (1,-1) ..controls (0.3,-0.8) and (-0.3,-0.8) .. (-1,-1);
\draw[-,thick] (-1,-1) to (-1,1);

\draw[-,dashed,thick,dash pattern= on 4pt off 8pt,dash phase=6pt,red] (-1,1) to (1,-1);
\draw[-,dashed,thick,dash pattern= on 4pt off 8pt,dash phase=6pt,red] (1,1) to (-1,-1);

\draw[-,dashed,thick,dash pattern= on 4pt off 8pt] (-1,1) to (1,-1);
\draw[-,dashed,thick,dash pattern= on 4pt off 8pt] (1,1) to (-1,-1);

\draw[-,thick] (1,1) .. controls (0.75,0.2) and (0.75,-0.2) .. (1,-1);
\draw[-,thick] (1,1) .. controls (0.4,0.2) and (0.4,-0.2) .. (1,-1);
\draw[-,thick] (1,1) .. controls (0.05,0.05) and (0.05,-0.05) .. (1,-1);

\draw[-,thick] (-1,1) .. controls (-0.75,0.2) and (-0.75,-0.2) .. (-1,-1);
\draw[-,thick] (-1,1) .. controls (-0.4,0.2) and (-0.4,-0.2) .. (-1,-1);
\draw[-,thick] (-1,1) .. controls (-0.05,0.05) and (-0.05,-0.05) .. (-1,-1);

\draw[-,thick,red] (-1,1) .. controls (-0.2,0.55) and (0.2,0.55) .. (1,1);
\draw[-,thick,red] (-1,1) .. controls (-0.2,0.3) and (0.2,0.3) .. (1,1);
\draw[-,thick,red] (-1,1) .. controls (-0.02,0.0175) and (0.02,0.0175) .. (1,1);

\draw[-,thick,red] (-1,-1) .. controls (-0.2,-0.55) and (0.2,-0.55) .. (1,-1);
\draw[-,thick,red] (-1,-1) .. controls (-0.2,-0.3) and (0.2,-0.3) .. (1,-1);
\draw[-,thick,red] (-1,-1) .. controls (-0.02,-0.0175) and (0.02,-0.0175) .. (1,-1);

\node[red] at (0,0) {$\bullet$};
\node at (0,0) {$\circ$};

\draw[->,very thick] (0.8,0) to (0.6,0) ;
\draw[<-,thin] (1.2,0.3) to (0.8,0.2);
\node at (1.8,0.38) {$z=z_0$ hypersurface};
\draw[->,very thick,red] (0,0.2) to (0,0.55);
\end{tikzpicture}
\caption{In this Penrose diagram of an eternal black hole, the right exterior is shaded in gray and the interior is shaded in red. The black arrow in the exterior shows an inwards pointing normal vector, which is space-like. Various slices shown by black lines are constant radial surfaces. The radial Hamiltonian evolves the slices in the direction of the normal vector. Similarly, in the interior, the red arrow shows the normal vector, which is time-like in nature. We have found the Hamiltonian that does the evolution for these time-like slices. The singularity is marked with zagged lines. In section \ref{interior}, we explore the implications of Hamiltonian constraints near the singularity.}%A two-sided asymptotically anti-de Sitter (AdS) black hole, with the exterior in gray and the interior in red. The lines are constant radial slices, with the horizon being the dashed lines. The black arrow represents a conventional UV $\to$ IR holographic RG flow while the red arrow indicates a trans-IR flow parameterized by a timelike radial coordinate.}
\label{figs:holoRGFlow}
\end{figure}
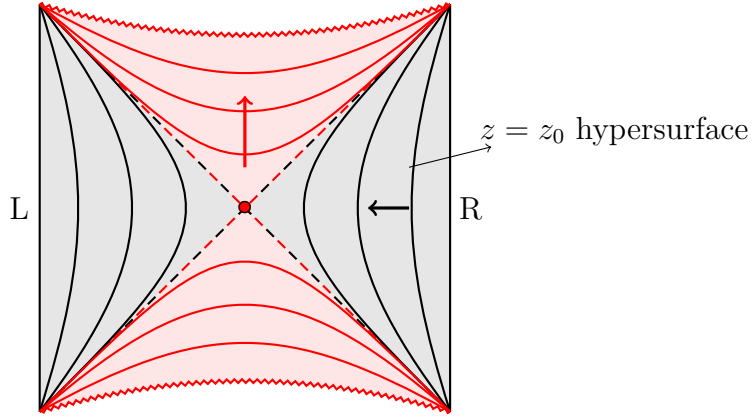

\section{Hamilton-Jacobi analysis at finite temperature}
\label{sec:thermal-hj-derivative-expansion}

In Sec.~\ref{superpotential section}, we described the Einstein equations
in terms of the blackening function $f$ and the superpotential $W(\phi)$.
We now formulate the same problem directly in Hamilton-Jacobi language.
This provides a finite-temperature extension of the holographic RG
framework developed in
refs.~\cite{deBoer:1999tgo,deBoer:2000cz,DeWolfe:1999cp,Kiritsis:2014kua}. To study it systematically, we
construct the Hamilton-Jacobi functional in a derivative expansion
along the boundary directions.\footnote{Throughout this section we set
\begin{equation}
    \frac{1}{16\pi G}=1.
\end{equation}
The radial coordinate is denoted by $u$, while the coordinates intrinsic
to a constant-$u$ slice are
\begin{equation}
    x^\mu=(\tau,x,y).
\end{equation}
}

At zeroth order in the derivative expansion, all fields depend only on
the radial coordinate $u$. There are no derivatives along the
constant-$u$ slices. Under our parity-even and covariant ansatz, there is no independent scalar contribution at first order in boundary derivatives. The first nontrivial derivative corrections therefore appear at second order. These include invariants such as
\begin{equation}
    R[\gamma],
    \qquad
    \gamma^{\mu\nu}\partial_\mu\phi\,\partial_\nu\phi,
\end{equation}
together with the corresponding two-derivative structures associated
with the finite-temperature geometry. The derivative expansion describes slowly varying deformations of the homogeneous black-brane background. We begin with the zeroth-order problem. We will also make the
connection with the first-order superpotential description
$(W,f)$ obtained in the previous section transparent.

\subsection{Hamilton-Jacobi problem}
\label{subsec:thermalHJsetup}
Hamilton-Jacobi (HJ) theory reformulates radial bulk evolution in terms of an on-shell action $S[\gamma,\phi]$ defined on a constant-radial slice. Its
functional derivatives give the canonical momenta, and therefore determine
the radial flow of the induced metric and matter fields. In holography, the
radial direction is identified with the renormalization-group scale, so the
HJ equation provides a natural framework for holographic RG flow.\\

We begin with the Euclidean Einstein-scalar action, including the
Gibbons-Hawking-York term,
\begin{equation}
S_E
=
-\int_{\mathcal M} d^4x\,\sqrt{g}
\left[
R-\frac12 g^{ab}\partial_a\phi\partial_b\phi-V(\phi)
\right]
-2\int_{\partial\mathcal M}d^3x\,\sqrt{\gamma}\,K
+S_{\rm ct},
\label{eq:HJfiniteAction}
\end{equation}
and choose proper-radial ADM gauge,
\begin{equation}
    ds^2=du^2+\gamma_{\mu\nu}(u,x)\,dx^\mu dx^\nu .
\end{equation}
The coordinate $u$ is taken to increase toward the interior.  We write
\begin{equation}
    \dot{}\equiv\partial_u,
    \qquad
    K_{\mu\nu}=\frac12\dot\gamma_{\mu\nu},
    \qquad
    K=\gamma^{\mu\nu}K_{\mu\nu}.
\label{eq:Kconvention}
\end{equation}
With these conventions, the Gauss-Codazzi decomposition of Lagrangian gives
\begin{equation}
S_E
=
-\int du\,d^3x\,\sqrt{\gamma}
\left[
K^2-K_{\mu\nu}K^{\mu\nu}
+R[\gamma]
-\frac12\dot\phi^2
-\frac12(\nabla\phi)^2
-V(\phi)
\right]
+S_{\rm ct},
\label{eq:HJreducedGeneral}
\end{equation}
where $R[\gamma]$ and $\nabla_\mu$ are constructed from the induced
metric and \footnote{
The orientation in \eqref{eq:Kconvention} is in the convention of inward pointing normal vector. The resulting sign change in
$K_{\mu\nu}$ is purely conventional. All equations in the present
section use \eqref{eq:Kconvention}.}
\begin{equation}
    (\nabla\phi)^2
    =\gamma^{\mu\nu}\partial_\mu\phi\,\partial_\nu\phi .
\end{equation}

The canonical momenta  are therefore
\begin{equation}
\Pi^{\mu\nu}
\equiv
\frac{1}{\sqrt{\gamma}}
\frac{\delta S}{\delta\gamma_{\mu\nu}}
=
K^{\mu\nu}-K\gamma^{\mu\nu},
\qquad
\Pi_\phi
\equiv
\frac{1}{\sqrt{\gamma}}
\frac{\delta S}{\delta\phi}
=
\dot\phi .
\label{eq:HJmomentaDef}
\end{equation}

The radial Hamiltonian constraint is
\begin{equation}
K_{\mu\nu}K^{\mu\nu}-K^2
+\frac12\dot\phi^2
+R[\gamma]
-\frac12(\nabla\phi)^2
-V(\phi)=0.
\label{eq:HJfiniteHam}
\end{equation}
Using
\begin{equation}
K=-\frac12\Pi,
\qquad
K_{\mu\nu}K^{\mu\nu}-K^2
=
\Pi_{\mu\nu}\Pi^{\mu\nu}
-\frac12\Pi^2,
\end{equation}
one obtains the Hamilton-Jacobi equation
\begin{align}
0={}&
\frac1{\gamma}
\bigg[
-\frac12
\left(
\gamma_{\mu\nu}
\frac{\delta S}{\delta\gamma_{\mu\nu}}
\right)^2
+
\gamma_{\mu\alpha}\gamma_{\nu\beta}
\frac{\delta S}{\delta\gamma_{\mu\nu}}
\frac{\delta S}{\delta\gamma_{\alpha\beta}}
+
\frac12
\left(\frac{\delta S}{\delta\phi}\right)^2
\bigg]
\nonumber\\
&\qquad
+R[\gamma]
-\frac12(\nabla\phi)^2
-V(\phi).
\label{eq:HJfiniteConstraint}
\end{align}
The radial momentum constraint is independently
\begin{equation}
-2\nabla_\mu
\left(
\frac{\delta S}{\delta\gamma_{\mu\nu}}
\right)
+
\frac{\delta S}{\delta\phi}\,
\nabla^\nu\phi
=0,
\label{eq:HJfiniteMomentumFunctional}
\end{equation}
or, equivalently,
\begin{equation}
\nabla_\mu\Pi^{\mu\nu}
=
\frac12\Pi_\phi\nabla^\nu\phi .
\label{eq:HJfiniteMomentum}
\end{equation}
Both equations are constraints on the Hamilton-Jacobi functional:
the Hamiltonian constraint controls radial evolution, whereas the
momentum constraint is the translational invariance of the constant-$u$ slices.

It is useful to introduce the bilinear bracket
\begin{align}
\{F,G\}
\equiv
\frac1{\sqrt{\gamma}}
\bigg[
&-\frac12
\left(
\gamma_{\mu\nu}\frac{\delta F}{\delta\gamma_{\mu\nu}}
\right)
\left(
\gamma_{\rho\sigma}\frac{\delta G}{\delta\gamma_{\rho\sigma}}
\right)
+
\gamma_{\mu\rho}\gamma_{\nu\sigma}
\frac{\delta F}{\delta\gamma_{\mu\nu}}
\frac{\delta G}{\delta\gamma_{\rho\sigma}}
\nonumber\\
&+
\frac12
\frac{\delta F}{\delta\phi}
\frac{\delta G}{\delta\phi}
\bigg].
\label{eq:HJbracketFinite}
\end{align}
Multiplying \eqref{eq:HJfiniteConstraint} by $\sqrt{\gamma}$ then gives
\begin{equation}
\{S,S\}
+\mathcal L^{(0)}
+\mathcal L^{(2)}
=0,
\qquad
\mathcal L^{(0)}
=-\sqrt{\gamma}\,V(\phi),
\qquad
\mathcal L^{(2)}
=
\sqrt{\gamma}
\left[
R[\gamma]
-\frac12(\nabla\phi)^2
\right].
\label{eq:HJlocalSources}
\end{equation}

As a normalization check, let
\begin{equation}
S^{(0)}
=
\int d^3x\,\sqrt{\gamma}\,U(\phi).
\label{check1}
\end{equation}
Then
\begin{equation}
\Pi^{\mu\nu}
=
\frac12 U\,\gamma^{\mu\nu},
\qquad
\Pi_\phi=U_\phi,
\end{equation}
These equations also gives the flow eq for the warp factor that we will discuss in the next section.
\begin{equation}
\{S^{(0)},S^{(0)}\}
=
\sqrt{\gamma}
\left(
\frac12U_\phi^2-\frac38U^2
\right).
\end{equation}
The zeroth-order Hamiltonian constraint equation consequently reproduces
\begin{equation}
V(\phi)
=
\frac12U_\phi^2-\frac38U^2,
\label{eq:zeroTsuperpotential}
\end{equation}
in agreement with Eq.~\eqref{superpotentialrelation}.

% ============================================================
\subsection{Hamilton-Jacobi at zeroth order: Homogeneous finite-temperature }
\label{sec:thermal-HJ-homogeneous}
% ============================================================
In the previous section we underscored the usefulness of the Hamilton-Jacobi formalism. 
%, we have setup the Hamilton Jacobi formalism. 
We now specialize to the  finite-temperature case. We organize the analysis  in a derivative expansion along the boundary directions, in analogy with the  fluid-gravity correspondence. The goal is to construct the Hamilton-Jacobi functional whose associated first-order flow, together with the Hamiltonian constraint, reproduces the bulk Einstein-scalar equations.
The functional $S^{0}$  constructed in  \eqref{check1} provides a simple example. Its variation yields the flow equation as well as relates the scalar potential to the superpotential. We first construct the homogeneous, zeroth-order Hamilton characteristic function and then extend the analysis to second order in derivatives.

%First, we find such functional at finite temperature but with zeroth order in derivative expansion and then generalize to the second order case.
  Consider the
Euclidean black-brane ansatz
\begin{equation}
ds_4^2
=
du^2+e^{2A(u)}
\left[
e^{2\sigma(u)}d\tau^2+dx^2+dy^2
\right],
\qquad
\phi=\phi(u),
\qquad
\tau\sim\tau+\beta .
\label{eq:HJhomogeneousAnsatz}
\end{equation}
The function $A$ controls the overall spatial scale, whereas $\sigma$
measures the relative size of the Euclidean thermal circle.  Equivalently,
the usual blackening function is
\begin{equation}
f=e^{2\sigma}.
\end{equation}
Finite temperature therefore introduces a relative temporal-spatial
difference that is absent in the zero-temperature domain-wall
geometry.

For the ansatz \eqref{eq:HJhomogeneousAnsatz},
\begin{equation}
\sqrt{\gamma}=e^{3A+\sigma},
\qquad
K^\tau{}_\tau=\dot A+\dot\sigma,
\qquad
K^i{}_j=\dot A\,\delta^i{}_j ,
\end{equation}
and the Einstein-scalar action \eqref{eq:HJreducedGeneral} reduces to
\begin{equation}
S_{\rm red}
=
\beta V_2
\int du\,e^{3A+\sigma}
\left[
-6\dot A^2
-4\dot A\dot\sigma
+\frac12\dot\phi^2
+V(\phi)
\right].
\label{eq:thermalReducedAction}
\end{equation}

There is one additional conserved quantity at finite temperature.  Let's define
\begin{equation}
\rho\equiv 3A+\sigma .
\end{equation}
Then
\begin{equation}
S_{\rm red}
=
\beta V_2
\int du\,e^\rho
\left[
-\frac23\dot\rho^{\,2}
+\frac23\dot\sigma^{\,2}
+\frac12\dot\phi^2
+V(\phi)
\right].
\end{equation}
In these variables $\sigma$ is cyclic.  Its conjugate momentum is therefore
radially conserved.  We parameterize this conserved momentum by
\begin{equation}
p_\sigma=-\frac23\beta V_2D .
\end{equation}
Hamilton's equation then gives
\begin{equation}
\dot\sigma
=
-\frac12D\,e^{-3A-\sigma},
\end{equation}
or equivalently
\begin{equation}
e^{3A+\sigma}\dot\sigma=-\frac{D}{2}.
\label{eq:thermalChargeConservation}
\end{equation}
In terms of $f=e^{2\sigma}$,
\begin{equation}
e^{3A-\sigma}\dot f=-D.
\label{eq:blackeningFromMomentum}
\end{equation}

This is the familiar blackening equation that we have obtained in \eqref{eq:D definition}. 
The integration constant $D$ is related to the temperature of the black brane solution as established in
Sec.~\ref{superpotential section},
\begin{equation}
D=Ts .
\label{eq:Dthermal}
\end{equation}
In a nutshell, the momentum conjugate to the thermal anisotropy is the product of
temperature and entropy density. From now on we return to the original variables $(A,\sigma,\phi)$. The
auxiliary variable $\rho$ will not be needed again.

% ------------------------------------------------------------
\subsubsection*{Thermal Hamilton-Jacobi function}
% ------------------------------------------------------------

The conserved charge $D$ suggests introducing the local combination
\begin{equation}
\eta
\equiv
D\,e^{-3A-\sigma}.
\label{eq:etaD}
\end{equation}
It measures the local strength of the thermal deformation along the radial
flow.  Near an asymptotically AdS boundary $\eta\rightarrow0$, even at
fixed nonzero $D$. The homogeneous HJ functional on a branch of fixed $D$
can be written as
\begin{equation}
S_D^{(0)}
=
\beta V_2
\left[
-\frac23D\sigma
+
\frac{D}{\eta}P(\phi,\eta)
\right].
\label{eq:S0D}
\end{equation}
The term linear in $\sigma$ implements the fixed momentum conjugate to the
thermal anisotropy.  The single function
\begin{equation}
P=P(\phi,\eta)
\end{equation}
is the finite-temperature generalization of the zero-temperature functional $U(\phi)$.

Substitution into the Hamiltonian constraint gives
\begin{equation}
V(\phi)
=
\frac12P_\phi^2
-\frac38
\left(P-\eta P_\eta\right)^2
+\frac16\eta^2 .
\label{eq:Pmaster}
\end{equation}
This is the homogeneous finite-temperature Hamilton-Jacobi equation.
Its ultraviolet boundary condition is
\begin{equation}
P(\phi,0)=U(\phi),
\label{eq:PzeroTlimit}
\end{equation}
where the ordinary zero-temperature superpotential satisfies
\begin{equation}
V(\phi)
=
\frac12U_\phi^2-\frac38U^2.
\end{equation}

It is useful from this point onward to define only two combinations,
\begin{align}
\lambda_s
&\equiv
\frac12\left(P-\eta P_\eta\right)
-\frac13\eta,
\label{eq:lambdaS}
\\
\lambda_\tau
&\equiv
\frac12\left(P-\eta P_\eta\right)
+\frac23\eta.
\label{eq:lambdaTau}
\end{align}
They have an immediate geometric interpretation:
\begin{equation}
\dot A=-\frac12\lambda_s,
\qquad
\dot A+\dot\sigma=-\frac12\lambda_\tau,
\qquad
\dot\phi=P_\phi .
\label{eq:homogeneousThermalFlows}
\end{equation}
The two quantities are therefore the radial expansion rates of the spatial
metric and of the thermal circle, respectively.  Their difference is
particularly simple,
\begin{equation}
\lambda_\tau-\lambda_s=\eta .
\label{eq:thermalAnisotropy}
\end{equation}
Equivalently,
\begin{equation}
\dot\sigma=-\frac12\eta .
\label{eq:sigmaFlowEta}
\end{equation}

Equation \eqref{eq:thermalAnisotropy} makes the finite-temperature physics
transparent.  At zero temperature the temporal and spatial directions
undergo the same radial Weyl rescaling.  At finite temperature the two
rates split, and the splitting is controlled directly by the local parameter $\eta$.

Since $D$ is radially conserved,
\begin{equation}
\dot\eta
=
\frac34\eta
\left(P-\eta P_\eta\right).
\label{eq:etaFlow}
\end{equation}
Therefore, for any coefficient $F(\phi,\eta)$ appearing later in the
derivative expansion,
\begin{equation}
\dot F
=
P_\phi F_\phi
+
\frac34\eta
\left(P-\eta P_\eta\right)F_\eta .
\label{eq:characteristicDerivative}
\end{equation}
Using this HJ functional at fixed temperature, we can find the flow eq and match it with Einstein eq written superpotential functional $W$ in section \ref{superpotential section}. The detailed map is established in appendix \ref{app:Wf-map}.

% ------------------------------------------------------------
\subsubsection*{Thermal RG flow}
% ------------------------------------------------------------

Taking $A$ as the logarithmic spatial RG scale,
\begin{equation}
\log\mu=A,
\end{equation}
the scalar beta function follows immediately:
\begin{equation}
\beta_\phi^{(A)}
\equiv
\frac{d\phi}{dA}
=
-\frac{2P_\phi}{\lambda_s}.
\label{eq:betaFiniteD}
\end{equation}
At $\eta=0$,
\begin{equation}
P\rightarrow U,
\qquad
\lambda_s,\lambda_\tau\rightarrow\frac12U,
\end{equation}
and hence
\begin{equation}
\beta_\phi^{(A)}
\longrightarrow
-\frac{4U_\phi}{U},
\end{equation}
as required.

The correction is not controlled directly by $T/\mu$.  Rather,
\begin{equation}
\eta=D\,e^{-3A-\sigma}.
\end{equation}
For an asymptotically conformal black brane in three boundary dimensions,
$D$ has scaling dimension three, so near the boundary
\begin{equation}
\eta\sim\left(\frac{T}{\mu}\right)^3 .
\end{equation}
Thus the thermal anisotropy is power-law suppressed in the ultraviolet, and the usual isotropic holographic RG flow is recovered.

% ------------------------------------------------------------
\subsubsection*{Canonical anisotropy and thermodynamics}
% ------------------------------------------------------------

The same splitting appears directly in the radial canonical momenta.  At
zeroth derivative order,
\begin{equation}
\Pi^\tau{}_\tau=\lambda_s,
\qquad
\Pi^i{}_j
=
\frac12(\lambda_s+\lambda_\tau)\delta^i{}_j,
\qquad
\Pi_\phi=P_\phi .
\label{eq:thermalHJmomentaSimple}
\end{equation}
Therefore
\begin{equation}
\Pi^x{}_x-\Pi^\tau{}_\tau
=
\frac{\eta}{2}.
\label{eq:canonicalThermalAnisotropy}
\end{equation}
The conserved charge $D$, through the local combination $\eta$ controls the anisotropy of the radial canonical stress response.

For the AdS-Schwarzschild case, after the usual local subtraction, the
boundary stress tensor is linear in $D$.  Its temporal and spatial
components obey
\begin{equation}
\epsilon=2p ,
\end{equation}
as required by the traceless equation of state of a $2+1$-dimensional
conformal theory.  Together with $D=Ts$, the Euler relation
\begin{equation}
\epsilon+p=Ts
\end{equation}
provides a consistency check between the horizon normalization and the
Brown-York stress tensor.  We will not repeat that here.
% ------------------------------------------------------------
\subsubsection*{What changes when the sources are inhomogeneous?}
% ------------------------------------------------------------

The homogeneous Hamiltonian constraint determines a family of radial
solutions labeled by $D$. Once the boundary sources are allowed
to depend on $x^i$, the momentum constraint must also be imposed.  At leading order in the derivative expansion, the conserved homogeneous charge may be promoted to slowly varying branch data \(D(x)\), while remaining radially conserved,
\begin{equation}
\partial_u D = 0 .
\end{equation}
Its spatial dependence is not arbitrary, however, and is constrained by the momentum constraint. Equivalently, it is convenient to describe the local thermal state through the scalar combination
\begin{equation}
\eta=\frac{D}{N\sqrt{h}},
\end{equation}
whose spatial variation is fixed by hydrostatic equilibrium.

Consequently, $S_D^{(0)}$ should be interpreted as the homogeneous Hamilton characteristic function on a fixed thermal branch.  When a derivative expansion is introduced, one must first vary the
Hamilton-Jacobi functional with respect to the independent sources and only afterwards impose the momentum constraint that selects the physical
hydrostatic branch.

This observation will be essential at second order.  It is also the
reason that the finite-temperature derivative expansion cannot be
obtained by simply taking the zero-temperature functional and replacing
the superpotential $U(\phi)$ by a temperature-dependent function.  The
thermal state carries an additional radial charge, and the momentum
constraint determines how that charge is allowed to respond to
inhomogeneous sources.

% ============================================================
\subsection{ Hamilton-Jacobi functional at two derivatives}
\label{sec:two-derivative}
% ============================================================

We now allow the thermal state and scalar source to vary slowly in space.
The homogeneous solution derived is the zeroth-order background. The spatial curvature, temperature gradients, scalar
gradients, etc. enter at second derivative order.

The natural stationary decomposition of the induced metric is
\begin{equation}
ds_3^2
=
N^2(d\tau+a_i dx^i)^2
+
h_{ij}dx^i dx^j ,
\label{eq:stationaryMetricNh}
\end{equation}
with all fields independent of $\tau$.  In terms of the variables used in
the homogeneous discussion,
\begin{equation}
N=e^{A+\sigma},
\qquad
h_{ij}=e^{2A}g_{ij}.
\label{eq:NhDefinitions}
\end{equation}
The quantity $N$ is the proper size of the Euclidean thermal circle, while
$h_{ij}$ is the physical spatial metric, $a_i$ is the KK gauge field. It has field strength
% The stationary shift transforms as
% \begin{equation}
% a_i\rightarrow a_i-\partial_i\Lambda
% \end{equation}
% under $\tau\rightarrow\tau+\Lambda(x)$, 

% so it enters through the invariant

\begin{equation}
\mathcal F_{ij}
=
\partial_i a_j-\partial_j a_i .
\end{equation}

At two spatial derivatives there are five independent parity-even local
structures:
\begin{equation}
R[h],
\qquad
N^2\mathcal F_{ij}\mathcal F^{ij},
\qquad
(\nabla\log N)^2,
\qquad
(\nabla\phi)^2,
\qquad
\nabla_i\log N\,\nabla^i\phi .
\label{eq:fivePhysicalStructures}
\end{equation}
The local Hamilton-Jacobi functional therefore has the general form
\begin{equation}
S_D^{(2)}
=
\beta
\int d^2x\,\sqrt h\,N
\left[
F_1R[h]
+
F_2N^2\mathcal F^2
+
F_3(\nabla\log N)^2
+
F_4(\nabla\phi)^2
+
F_5\nabla\log N\cdot\nabla\phi
\right],
\label{eq:S2Nh}
\end{equation}
where
\begin{equation}
F_n=F_n(\phi,\eta),
\qquad n=1,\ldots,5.
\end{equation}

The five functions have a direct physical interpretation:

\begin{itemize}
\item $F_1$ controls the response to spatial curvature;
\item $F_2$ controls the stationary Kaluza-Klein or vorticity sector;
\item $F_3$ controls gradients of the local thermal-circle size;
\item $F_4$ controls scalar gradients;
\item $F_5$ mixes scalar and temperature gradients.
\end{itemize}

The temperature parameter \(D\) is held fixed when taking functional derivatives of the Hamilton-Jacobi functional. After the canonical responses have been obtained, \(D\) may be promoted to slowly varying local parameter, whose allowed spatial dependence is then determined by the momentum constraint. This ordering is important: the Hamilton-Jacobi functional is first varied with respect to the independent sources at fixed parameter, while hydrostatic equilibrium is imposed subsequently as a constraint on the resulting canonical data. This is also the reason that we don't have these terms $(\nabla D)^2, \nabla D . \nabla \phi, \, \nabla D . \log N$ in the second order action ansatz \eqref{eq:S2Nh}.

% ------------------------------------------------------------
\subsubsection*{Canonical responses}
% ------------------------------------------------------------

It is convenient to denote the responses of the two-derivative functional
to the four stationary sources by
\begin{equation}
\mathcal E
\equiv
\frac{1}{\beta\sqrt h\,N}
\frac{\delta S_D^{(2)}}{\delta\log N},
\qquad
\mathcal T^{ij}
\equiv
\frac{2}{\beta\sqrt h\,N}
\frac{\delta S_D^{(2)}}{\delta h_{ij}},
\label{eq:ETresponsesSimple}
\end{equation}
and
\begin{equation}
\mathcal J^i
\equiv
\frac{1}{\beta\sqrt h\,N}
\frac{\delta S_D^{(2)}}{\delta a_i},
\qquad
\mathcal O
\equiv
\frac{1}{\beta\sqrt h\,N}
\frac{\delta S_D^{(2)}}{\delta\phi}.
\label{eq:JOresponsesSimple}
\end{equation}
These are respectively the lapse, spatial stress, current, and
scalar responses.  They are all of second derivative order.

On the fixed-$D$ branch, the local thermal variable obeys
\begin{equation}
\delta \log \eta
=
-\delta \log N
-\frac{1}{2}h^{ij}\delta h_{ij},
\label{eq:etaVariationFixedD}
\end{equation}
 Consequently,
the coefficient functions $F_n(\phi,\eta)$ also vary under lapse and
spatial-metric variations.  The explicit lapse response is
\begin{align}
\mathcal E={}&
(F_1-\eta F_{1,\eta})R[h]
+
(3F_2-\eta F_{2,\eta})N^2\mathcal F^2
\nonumber\\
&+
(F_3-\eta F_{3,\eta})(\nabla\log N)^2
+
(F_4-\eta F_{4,\eta})(\nabla\phi)^2
\nonumber\\
&+
(F_5-\eta F_{5,\eta})
\nabla\log N\cdot\nabla\phi
\nonumber\\
&-
\frac{2}{N}
\nabla_i\!\left(
NF_3\nabla^i\log N
\right)
-
\frac{1}{N}
\nabla_i\!\left(
NF_5\nabla^i\phi
\right).
\label{eq:lapseResponseSecondSimple}
\end{align}

The trace of the spatial stress is
\begin{align}
\mathcal T
\equiv h_{ij}\mathcal T^{ij}
={}&
-\frac{2}{N}\nabla^2(NF_1)
-2\eta F_{1,\eta}R[h]
\nonumber\\
&-
2N^2(F_2+\eta F_{2,\eta})\mathcal F^2
-2\eta F_{3,\eta}(\nabla\log N)^2
\nonumber\\
&-
2\eta F_{4,\eta}(\nabla\phi)^2
-2\eta F_{5,\eta}
\nabla\log N\cdot\nabla\phi .
\label{eq:traceResponseSecondSimple}
\end{align}
Its traceless part follows from
\begin{align}
\mathcal T^{ij}={}&
\frac{2}{N}
\left[
\nabla^i\nabla^j(NF_1)
-h^{ij}\nabla^2(NF_1)
\right]
-\eta F_{1,\eta}R[h]h^{ij}
\nonumber\\
&+
N^2
\left[
(F_2-\eta F_{2,\eta})\mathcal F^2h^{ij}
-4F_2\mathcal F^i{}_k\mathcal F^{jk}
\right]
\nonumber\\
&+
(F_3-\eta F_{3,\eta})(\nabla\log N)^2h^{ij}
-2F_3\nabla^i\log N\nabla^j\log N
\nonumber\\
&+
(F_4-\eta F_{4,\eta})(\nabla\phi)^2h^{ij}
-2F_4\nabla^i\phi\nabla^j\phi
\nonumber\\
&+
(F_5-\eta F_{5,\eta})
(\nabla\log N\cdot\nabla\phi)h^{ij}
-2F_5\nabla^{(i}\log N\nabla^{j)}\phi .
\label{eq:metricResponseSecondSimple}
\end{align}
In two spatial dimensions the $\mathcal F^2$ contribution is purely trace.

The remaining two responses are
\begin{equation}
\mathcal J^i
=
-\frac4N
\nabla_j
\left(
N^3F_2\mathcal F^{ji}
\right),
\label{eq:KKresponseSecondSimple}
\end{equation}
and
\begin{align}
\mathcal O={}&
F_{1,\phi}R[h]
+
F_{2,\phi}N^2\mathcal F^2
+
F_{3,\phi}(\nabla\log N)^2
+
F_{4,\phi}(\nabla\phi)^2
\nonumber\\
&+
F_{5,\phi}\nabla\log N\cdot\nabla\phi
-
\frac1N\nabla_i
\left[
N
\left(
2F_4\nabla^i\phi
+
F_5\nabla^i\log N
\right)
\right].
\label{eq:scalarResponseSecondSimple}
\end{align}

These expressions contain all information needed both for the
Hamiltonian constraint and for the radial characteristic flow.

% ------------------------------------------------------------
\subsubsection*{The second-order Hamiltonian constraint}
% ------------------------------------------------------------

The Einstein-scalar Hamiltonian constraint is
\begin{equation}
R[\gamma]
-\frac12(D\phi)^2
+
K_{\mu\nu}K^{\mu\nu}
-K^2
+
\frac12\dot\phi^2
-
V(\phi)
=0.
\label{eq:HamiltonianConstraintGeometric}
\end{equation}
Its zeroth-order part is exactly the homogeneous thermal
Hamilton-Jacobi equation \eqref{eq:Pmaster}.  At second-derivative order
the constraint reduces to the compact canonical identity \footnote{This can be explicitly shown by computing all the terms in the Hamiltonian constraints and identifying the momenta as
\begin{equation}
\Pi^\tau{}_{\tau(0)}
=
\lambda_s,
\qquad
\Pi^i{}_{j(0)}
=
\frac12\left(\lambda_s+\lambda_\tau\right)\delta^i{}_j .
\label{eq:homogeneousCanonicalMomenta}
\end{equation}
Their trace is therefore
\begin{equation}
\Pi_{(0)}
\equiv
\Pi^\mu{}_{\mu(0)}
=
2\lambda_s+\lambda_\tau .
\label{eq:homogeneousMomentumTrace}
\end{equation}

For the two-derivative functional, the response definitions
\eqref{eq:ETresponsesSimple}--\eqref{eq:JOresponsesSimple} imply
\begin{equation}
\Pi^\tau{}_{\tau(2)}
=
\frac12\mathcal E,
\qquad
\Pi^i{}_{i(2)}
=
\frac12\mathcal T,
\qquad
\Pi_{\phi(2)}
=
\mathcal O .
\label{eq:secondOrderCanonicalMomenta}
\end{equation}
For the stationary Kaluza-Klein metric,
\begin{equation}
ds_3^2
=
N^2(d\tau+a)^2
+
h_{ij}dx^i dx^j ,
\end{equation}
the three-dimensional Ricci scalar decomposes as
\begin{equation}
R[\gamma]
=
R[h]
-\frac14 N^2 \mathcal F_{ij}\mathcal F^{ij}
-\frac{2}{N}\nabla^2 N .
\label{eq:KKRicciDecomposition}
\end{equation}}
\begin{equation}
0=
\int d^2x\sqrt h\,N
\left[
-\frac12\lambda_\tau\mathcal E
-\frac12\lambda_s\mathcal T
+
P_\phi\mathcal O
+
R[h]
-\frac14N^2\mathcal F^2
-\frac12(\nabla\phi)^2
\right].
\label{eq:HJ2CanonicalResponseForm}
\end{equation}
This equation has a simple interpretation.  The homogeneous radial
background acts on the second-order functional through the temporal rate
$\lambda_\tau$, the spatial rate $\lambda_s$, and the scalar sector
$P_\phi$.  The remaining three terms are the intrinsic two-derivative
terms already present in the bulk Hamiltonian constraint.

This is the useful starting point for the descent equations.  In
particular, all integrations by parts are performed only after the
functional variation.  This keeps the derivatives of
$\lambda_\tau$, $\lambda_s$, and $P_\phi$ that are required for the correct
zero-temperature limit.

% ------------------------------------------------------------
\subsubsection*{Momentum constraint and hydrostatic equilibrium}
% ------------------------------------------------------------

The Hamiltonian constraint determines radial evolution, but it does not specify the spatial variation. That information comes from the momentum constraint,
\begin{equation}
\nabla_\nu K^\nu{}_i-\nabla_iK
=
\frac12\dot\phi\,\nabla_i\phi .
\label{eq:momentumConstraint}
\end{equation}

We now choose a reference radial slice on which
\begin{equation}
\nabla_iA=0 .
\label{eq:referenceSliceA}
\end{equation}
This choice is made only after the functional variation.  On this slice
the leading momentum constraint gives
\begin{equation}
\nabla_i\log\eta
=
\frac{3}
{1+3\partial_\eta(P-\eta P_\eta)}
\left(
P_{\phi\eta}\nabla_i\phi
-
\nabla_i\sigma
\right).
\label{eq:gradEtaHydrostaticSimple}
\end{equation}
Thus $\eta$, or equivalently the local thermal branch data $D(x)$, is not
an independent arbitrary function on a hydrostatic solution.  Its
spatial variation is fixed by the scalar and thermal-lapse gradients.

For later use, define
\begin{equation}
\Xi
\equiv
1+3\partial_\eta(P-\eta P_\eta).
\label{eq:XiSimple}
\end{equation}
Then for any function $G(\phi,\eta)$,
\begin{equation}
\nabla_iG
=
G_{|\phi}\nabla_i\phi
+
G_{|\sigma}\nabla_i\sigma ,
\label{eq:hydroChainSimple}
\end{equation}
where
\begin{equation}
G_{|\phi}
\equiv
G_\phi
+
\frac{3\eta P_{\phi\eta}}{\Xi}G_\eta,
\qquad
G_{|\sigma}
\equiv
-\frac{3\eta}{\Xi}G_\eta .
\label{eq:hydroDerivativesSimple}
\end{equation}
The vertical bar here simply means
``differentiate after restricting to the hydrostatic branch.''

% ------------------------------------------------------------
\subsubsection*{Five finite-temperature descent equations}
% ------------------------------------------------------------
Using the Hamiltonian constraint \eqref{eq:HJ2CanonicalResponseForm}, momentum constraint to restrict to hydrostatic branch and using integration by parts gives the the integrated constraint in terms of these 5 structures. 
Matching the five structures in \eqref{eq:fivePhysicalStructures} gives a
closed linear system for the five coefficient functions.  We use
\begin{equation}
\dot F_n
=
P_\phi F_{n,\phi}
+
\frac34\eta
\left(P-\eta P_\eta\right)
F_{n,\eta}
\end{equation}
for differentiation along the homogeneous radial characteristic.

The curvature coefficient obeys
\begin{equation}
\dot F_1
-\frac12\lambda_\tau F_1
+1=0 .
\label{eq:F1thermalSimple}
\end{equation}

The stationary Kaluza--Klein coefficient obeys
\begin{equation}
\dot F_2
-
\left(
\frac12\lambda_\tau+\eta
\right)F_2
-\frac14
=0 .
\label{eq:F2thermalSimple}
\end{equation}

The coefficient of the local-temperature gradient satisfies
\begin{equation}
0=
\dot F_3
-
\left[
\frac12\lambda_\tau
+
(\lambda_\tau)_{|\sigma}
\right]F_3
+
(P_\phi)_{|\sigma}F_5
-
\left(
F_1+(F_1)_{|\sigma}
\right)
(\lambda_s)_{|\sigma}.
\label{eq:F3thermalSimple}
\end{equation}

The scalar-gradient coefficient satisfies
\begin{equation}
0=
\dot F_4
+
\left[
2(P_\phi)_{|\phi}
-\frac12\lambda_\tau
\right]F_4
-
\frac12(\lambda_\tau)_{|\phi}F_5
-
(F_1)_{|\phi}(\lambda_s)_{|\phi}
-\frac12 .
\label{eq:F4thermalSimple}
\end{equation}

Finally, the mixed scalar--temperature coefficient obeys
\begin{align}
0={}&
\dot F_5
+
\left[
(P_\phi)_{|\phi}
-\frac12(\lambda_\tau)_{|\sigma}
-\frac12\lambda_\tau
\right]F_5
-
(\lambda_\tau)_{|\phi}F_3
\nonumber\\
&+
2(P_\phi)_{|\sigma}F_4
-
\left[
(F_1)_{|\phi}(\lambda_s)_{|\sigma}
+
\left(
F_1+(F_1)_{|\sigma}
\right)
(\lambda_s)_{|\phi}
\right].
\label{eq:F5thermalSimple}
\end{align}

Equations
\eqref{eq:F1thermalSimple}--\eqref{eq:F5thermalSimple}
are the complete two-derivative finite-temperature Hamilton-Jacobi
descent equations.  Their structure is physically transparent.  $F_1$
and $F_2$ are transported independently along the homogeneous thermal
flow, whereas the gradient sectors $F_3,F_4,F_5$ mix because gradients of
the scalar and of the local thermal circle are linked by hydrostatic
equilibrium.

% ------------------------------------------------------------
\subsubsection*{Radial flow through two derivatives}
% ------------------------------------------------------------

The same functional that solves the Hamiltonian constraint also generates
the radial flow.  Canonical inversion gives \footnote{The characteristic equations generated by the Hamilton-Jacobi functional
are equivalent to the radial Einstein-scalar equations.  Indeed, the
canonical momentum obtained directly from the bulk action obeys
\begin{equation}
\Pi^\mu{}_\nu
=
K^\mu{}_\nu-K\delta^\mu{}_\nu ,
\end{equation}
which in three induced dimensions is inverted as
\begin{equation}
K^\mu{}_\nu
=
\Pi^\mu{}_\nu
-\frac12\Pi\,\delta^\mu{}_\nu .
\end{equation}
Using $K_{\mu\nu}=\frac12\dot\gamma_{\mu\nu}$ together with the
Hamilton--Jacobi identification
$\pi^{\mu\nu}=\delta S/\delta\gamma_{\mu\nu}$ and
$\pi_\phi=\delta S/\delta\phi$ reproduces the radial flow equations
derived above.  Thus the Hamilton-Jacobi flow is simply the first-order
canonical form of the radial Einstein-scalar evolution.}
\begin{equation}
\frac{\dot N}{N}
=
-\frac12\lambda_\tau
+
\frac14
\left(
\mathcal E-\mathcal T
\right)
+
O(\partial^4),
\label{eq:NflowSecondSimple}
\end{equation}
\begin{equation}
h^{ik}\dot h_{kj}
=
-\lambda_s\delta^i{}_j
+
\mathcal T^i{}_j
-
\frac12
\left(
\mathcal E+\mathcal T
\right)\delta^i{}_j
+
O(\partial^4),
\label{eq:hflowSecondSimple}
\end{equation}
together with
\begin{equation}
\dot a_i
=
\frac{1}{N^2}\mathcal J_i
+
O(\partial^4),
\qquad
\dot\phi
=
P_\phi+\mathcal O
+
O(\partial^4).
\label{eq:aphiFlowSecondSimple}
\end{equation}

These equations are already the complete stationary characteristic flow
in terms of the physical sources $(N,h_{ij},a_i,\phi)$. To connect more directly with the homogeneous variables, write
\begin{equation}
N=e^{A+\sigma},
\qquad
h_{ij}=e^{2A}g_{ij},
\qquad
g^{ij}\dot g_{ij}=0.
\end{equation}
Then
\begin{align}
\dot A
&=
-\frac12\lambda_s
-\frac14\mathcal E
+
O(\partial^4),
\label{eq:AflowSecondSimple}
\\
\dot\sigma
&=
-\frac12\eta
+
\frac12\mathcal E
-\frac14\mathcal T
+
O(\partial^4),
\label{eq:sigmaFlowSecondSimple}
\\
g^{ik}\dot g_{kj}
&=
\mathcal T^i{}_j
-\frac12\mathcal T\,\delta^i{}_j
+
O(\partial^4).
\label{eq:gflowSecondSimple}
\end{align}

% ------------------------------------------------------------
\subsubsection*{Zero-temperature Limit}
% ------------------------------------------------------------

A strong consistency check is obtained by taking the zero-temperature
limit of the above construction. In this limit
\begin{equation}
\eta\rightarrow0,
\qquad
P(\phi,\eta)\rightarrow U(\phi),
\qquad
\lambda_s,\lambda_\tau\rightarrow\frac12U(\phi).
\label{eq:zeroTHomogeneousSimple}
\end{equation}

The apparently distinct temperature-gradient, gauge field field strength, curvature, and mixed-gradient structures are not independent
when the thermal anisotropy disappears.  They recombine precisely into
the ordinary covariant Einstein-scalar functional. The five stationary structures must then recombine into the covariant
three-dimensional functional
\begin{equation}
S_{\eta=0}^{(2)}
=
\int d^3x\sqrt\gamma
\left[
F_1(\phi)R[\gamma]
+
F_4(\phi)
D_\mu\phi D^\mu\phi
\right].
\label{eq:S2CovariantZeroT}
\end{equation}
This covariance requirement fixes
\begin{equation}
F_2=-\frac14F_1,
\qquad
F_3=0,
\qquad
F_5=2F_{1,\phi},
\label{eq:geometricZeroTRelationsSimple}
\end{equation}
while
\begin{equation}
F_4
=
\frac{U}{2U_\phi}F_{1,\phi}.
\label{eq:F4zeroTSimple}
\end{equation}
The only independent curvature equation is
\begin{equation}
U_\phi F_{1,\phi}
-\frac14UF_1
+1
=0 .
\label{eq:F1zeroTSimple}
\end{equation}
Once this equation and the geometric relations above are imposed, all five
finite-temperature descent equations reduce consistently to the single
covariant zero-temperature Hamilton-Jacobi equation. More details are presented in the appendix \ref{app:zeroT-five-equations}. This matches with the de-Boer et.al. \cite{deBoer:1999tgo,deBoer:2000cz,fukuma2000weyl} relations at two derivative order.

The corresponding zero-temperature characteristic equations are
\begin{align}
\dot\gamma_{\mu\nu}
={}&
-\frac12U\gamma_{\mu\nu}
-2F_1R_{\mu\nu}
+\frac12F_1R\,\gamma_{\mu\nu}
+2D_\mu D_\nu F_1
\nonumber\\
&+
\frac12F_4(D\phi)^2\gamma_{\mu\nu}
-2F_4D_\mu\phi D_\nu\phi
+
O(\partial^4),
\label{eq:covariantMetricFlowZeroT}
\end{align}
and
\begin{equation}
\dot\phi
=
U_\phi
+
F_{1,\phi}R
-
2F_4D^2\phi
-
F_{4,\phi}(D\phi)^2
+
O(\partial^4).
\label{eq:covariantScalarFlowZeroT}
\end{equation}
These equations provide a second zero-temperature consistency check, now at the
level of the radial characteristic flow rather than the descent equations.
After taking $\eta\rightarrow0$, the finite-temperature flows for
$(N,h_{ij},a_i,\phi)$ must recombine into the manifestly three-dimensional
covariant evolution equations for $(\gamma_{\mu\nu},\phi)$.  This confirms
that both the HJ coefficients and the resulting radial dynamics recover the
standard zero-temperature covariant theory.

The order of operations matters: the $\eta\rightarrow0$ limit is taken
after varying the finite-temperature functional.  Taking the limit inside
intermediate hydrostatic relations can incorrectly discard derivatives
that remain finite and are required for the covariant formalism.

% ============================================================
\subsection{Thermal holographic RG as an anisotropic Weyl flow}
\label{sec:thermal-radial-rg}
% ============================================================
The Hamilton-Jacobi equations admit a direct
renormalization-group interpretation.  For the stationary sources
\begin{equation}
h_{ij},\qquad N,\qquad a_i,\qquad \phi ,
\end{equation}
their radial velocities define the corresponding beta-functionals,
\begin{equation}
\beta^h_{ij}=\dot h_{ij},
\qquad
\beta_N=\dot N,
\qquad
\beta_i^a=\dot a_i,
\qquad
\beta_\phi=\dot\phi .
\end{equation}
The radial RG operator is therefore
\begin{equation}
\mathcal D_D
=
\int d^2x
\left[
\beta^h_{ij}\frac{\delta}{\delta h_{ij}}
+
\beta_N\frac{\delta}{\delta N}
+
\beta_i^a\frac{\delta}{\delta a_i}
+
\beta_\phi\frac{\delta}{\delta\phi}
\right].
\label{eq:fullRGoperatorSimple}
\end{equation}
If the Hamilton-Jacobi functional is separated into its local part and
a finite remainder $\Gamma_D$, the finite HJ equation takes the
Callan-Symanzik form
\begin{equation}
\mathcal D_D\Gamma_D=\mathcal A_D ,
\label{eq:GammaCSSimple}
\end{equation}
where $\mathcal A_D$ is a possible anomaly. In 3d boundaries, there are no gravitational contributions to anomalies. Furthermore, we can separate
the spatial beta-functional into its trace and traceless parts,
\begin{equation}
\beta^h_{ij}
=
2\omega\,h_{ij}
+
\widehat\beta^h_{ij},
\qquad
h^{ij}\widehat\beta^h_{ij}=0 ,
\label{eq:spatialFlowDecomposition}
\end{equation}
with
\begin{equation}
\omega
=
\frac14 h^{ij}\beta^h_{ij}.
\label{eq:omegaDefinition}
\end{equation}
We also define the relative running of the thermal lapse by
\begin{equation}
\chi
=
\frac{\beta_N}{N}-\omega .
\label{eq:chiDefinition}
\end{equation}
Thus $\omega$ measures the common Weyl rescaling of the stationary
geometry, while $\chi$ measures the evolution of the thermal circle
relative to the spatial metric.

Using the two-derivative characteristic equations derived above gives
\begin{align}
\omega
&=
-\frac12\lambda_s
-\frac14\mathcal E
+O(\partial^4),
\label{eq:omegaHJ}
\\
\chi
&=
-\frac12(\lambda_\tau-\lambda_s)
+\frac12\mathcal E
-\frac14\mathcal T
+O(\partial^4)
\nonumber\\
&=
-\frac12\eta
+\frac12\mathcal E
-\frac14\mathcal T
+O(\partial^4),
\label{eq:chiHJ}
\end{align}
while the volume-preserving spatial deformation is
\begin{equation}
\widehat\beta^{h\,i}{}_{j}
=
\mathcal T^i{}_j
-\frac12\mathcal T\,\delta^i{}_j
+O(\partial^4).
\label{eq:tracelessSpatialBeta}
\end{equation}

Equivalently, writing
\begin{equation}
N=e^{A+\sigma},
\qquad
h_{ij}=e^{2A}g_{ij},
\qquad
g^{ij}\dot g_{ij}=0 ,
\end{equation}
one has
\begin{equation}
\omega=\dot A,
\qquad
\chi=\dot\sigma,
\qquad
\widehat\beta^{h\,i}{}_{j}=g^{ik}\dot g_{kj}.
\label{eq:geometricFlowIdentification}
\end{equation}
The radial flow therefore separates naturally into three geometric
pieces: an overall Weyl rescaling $A$, a relative thermal-circle
deformation $\sigma$, and a traceless spatial deformation carried by
$g_{ij}$. Indeed, introducing the stationary Weyl generator
\begin{equation}
\mathcal W
=
\int d^2x
\left(
2h_{ij}\frac{\delta}{\delta h_{ij}}
+
N\frac{\delta}{\delta N}
\right),
\label{eq:stationaryWeylGenerator}
\end{equation}
the full RG operator can be written as
\begin{equation}
\begin{aligned}
\mathcal D_D
={}&
\int d^2x\,\omega
\left(
2h_{ij}\frac{\delta}{\delta h_{ij}}
+
N\frac{\delta}{\delta N}
\right)
\\
&+
\int d^2x
\left[
\chi N\frac{\delta}{\delta N}
+
\widehat\beta^h_{ij}
\frac{\delta}{\delta h_{ij}}
+
\beta_i^a\frac{\delta}{\delta a_i}
+
\beta_\phi\frac{\delta}{\delta\phi}
\right].
\end{aligned}
\label{eq:RGGeometricDecomposition}
\end{equation}
The first line is the ordinary Weyl part of the radial evolution.
The second contains the departures from a pure Weyl transformation:
relative thermal-circle running, traceless spatial deformation,
Kaluza--Klein flow, and scalar running.

For a homogeneous thermal state all derivative corrections vanish.
The gravitational flow then reduces to
\begin{equation}
\omega^{(0)}
=
-\frac12\lambda_s,
\qquad
\chi^{(0)}
=
-\frac12\eta,
\qquad
\widehat\beta^h_{ij}=0,
\qquad
\beta_i^a=0 .
\label{eq:homogeneousGeometricFlow}
\end{equation}
Consequently,
\begin{equation}
\mathcal D_D^{(0)}
=
\int d^2x
\left[
-\lambda_s h_{ij}\frac{\delta}{\delta h_{ij}}
-\frac12\lambda_\tau
N\frac{\delta}{\delta N}
+
P_\phi\frac{\delta}{\delta\phi}
\right].
\label{eq:RG0operatorSimple}
\end{equation}
The distinction between the temporal and spatial eigenvalues is
therefore precisely the finite-temperature deformation:
\begin{equation}
\lambda_\tau-\lambda_s=\eta .
\end{equation}
When $\eta\neq0$, the proper thermal circle and the spatial geometry
run at different rates, so radial evolution cannot be identified with
a single Weyl transformation.

Near the asymptotic boundary,
\begin{equation}
\eta\rightarrow0,
\qquad
\lambda_\tau-\lambda_s\rightarrow0 ,
\end{equation}
and this anisotropy disappears.  The gravitational part of the radial
flow then approaches an ordinary boundary Weyl transformation.
Upon rewriting this Weyl transformation as a dilation of the boundary
coordinates, the Euclidean period rescales together with the insertion
points.  The corresponding thermal Callan-Symanzik equation is
therefore
\begin{equation}
\left[
\mathscr D_x
+
\beta\frac{\partial}{\partial\beta}
+
\beta_{\phi_0}^{R}
\frac{\partial}{\partial\phi_0}
\right]
\left\langle
\mathcal O_1^R(x_1)\cdots
\mathcal O_n^R(x_n)
\right\rangle_\beta
+
\sum_{k=1}^n\gamma_k^R
\left\langle
\mathcal O_1^R(x_1)\cdots
\mathcal O_n^R(x_n)
\right\rangle_\beta
=0 .
\label{eq:thermalCSBoundarySimple}
\end{equation}
In particular, the term $\beta\partial_\beta$ does not represent an
independent Weyl transformation: it appears because a boundary dilation
rescales both physical distances and the thermal period.

% ============================================================
\subsection{AdS-Schwarzschild example}
\label{app:AdS-Schwarzschild-check}
% ============================================================

The planar AdS-Schwarzschild solution provides a direct check of the
finite-temperature Hamilton-Jacobi construction.  At an extremum of
the scalar potential,
\begin{equation}
\phi=\phi_0,
\qquad
V(\phi_0)=-\frac6{\ell^2},
\qquad
P_\phi(\phi_0,\eta)=0,
\end{equation}
the homogeneous HJ equation reduces to
\begin{equation}
-\frac6{\ell^2}
=
-\frac38
\left(P-\eta P_\eta\right)^2
+\frac16\eta^2 .
\label{app:eq:AdSBH-homogeneous}
\end{equation}
The branch continuously connected to pure AdS is
\begin{equation}
P_{\rm AdS BH}(\eta)
=
\sqrt{
\frac{16}{\ell^2}
+
\frac{4\eta^2}{9}
}
-
\frac23\eta\,
\operatorname{arcsinh}
\left(
\frac{\ell\eta}{6}
\right).
\label{app:eq:P-AdSBH}
\end{equation}
This is the homogeneous master function entering the zeroth-order Hamilton-Jacobi characteristic function for the black-brane.

It is convenient to introduce
\begin{equation}
a\equiv\frac{\ell\eta}{6},
\qquad
w\equiv\sqrt{1+a^2}.
\end{equation}
Then
\begin{equation}
P-\eta P_\eta=\frac4\ell w,
\qquad
\lambda_\tau=\frac2\ell(w+2a),
\qquad
\lambda_s=\frac2\ell(w-a),
\label{app:eq:AdSBH-lambdas}
\end{equation}
so that
\begin{equation}
\lambda_\tau-\lambda_s
=
\frac{6a}{\ell}
=
\eta ,
\end{equation}
as required by the general thermal characteristic equations.  The radial
derivative becomes
\begin{equation}
\frac{d}{du}
=
\frac{3aw}{\ell}\frac{d}{da}.
\label{app:eq:AdSBH-radial-derivative}
\end{equation}

The small-$\eta$ expansion,
\begin{equation}
P_{\rm AdS BH}
=
\frac4\ell
-\frac{\ell}{18}\eta^2
+O(\eta^4),
\end{equation}
makes explicit that the thermal branch approaches the pure-AdS
Hamilton-Jacobi solution smoothly as $\eta\rightarrow0$.

% ------------------------------------------------------------
\subsubsection*{Two-derivative HJ functional}
% ------------------------------------------------------------
At second order, we solve equations
\eqref{eq:F1thermalSimple}--\eqref{eq:F5thermalSimple} for the coefficients $F_n$. Once we have $P_{\rm AdS BH}$, we can solve those equations for our case. 
The curvature equation can be solved exactly and gives
\begin{equation}
F_1^{\rm AdS BH}
=
\ell(w+a).
\label{app:eq:F1-AdSBH}
\end{equation}
The Kaluza-Klein coefficient obeys
\begin{equation}
3aw\frac{dF_2}{da}
-
(w+8a)F_2
-
\frac{\ell}{4}
=
0,
\label{app:eq:F2-AdSBH}
\end{equation}
whose UV analytic solution begins as
\begin{equation}
F_2
=
-\frac{\ell}{4}
-\ell a
-\frac{13\ell}{8}a^2
-\frac{3\ell}{2}a^3
-\frac{23\ell}{32}a^4
+O(a^5).
\label{app:eq:F2-AdSBH-series}
\end{equation}

For the temperature-gradient sector one finds ($Q=P- \eta P_{\eta}$)
\begin{equation}
1+3Q_\eta=\frac{w+2a}{w},
\qquad
\partial_\sigma^{\rm h}
=
-\frac{3aw}{w+2a}\frac{d}{da}.
\label{app:eq:AdSBH-hydrostatic-derivative}
\end{equation}
In particular,
\begin{equation}
\partial_\sigma^{\rm h}\lambda_\tau
=
-\frac{6a(a+2w)}
{\ell(w+2a)},
\qquad
\partial_\sigma^{\rm h}\lambda_s
=
\frac{6a(w-a)}
{\ell(w+2a)},
\end{equation}
and the exact curvature solution satisfies
\begin{equation}
F_1+\partial_\sigma^{\rm h}F_1
=
\frac{\ell}{w+2a}.
\end{equation}
The $F_3$ equation therefore reduces to
\begin{equation}
3aw\frac{dF_3}{da}
-
\frac{1-a^2-8aw}{w+2a}F_3
-
\frac{6a\ell(w-a)}
{(w+2a)^2}
=
0,
\label{app:eq:F3-AdSBH}
\end{equation}
with UV-regular solution
\begin{equation}
F_3
=
\ell
\left[
3a-12a^2+33a^3-78a^4+O(a^5)
\right].
\label{app:eq:F3-AdSBH-series}
\end{equation}

The three coefficients therefore approach
\begin{equation}
F_1\rightarrow\ell,
\qquad
F_2\rightarrow-\frac{\ell}{4},
\qquad
F_3\rightarrow0
\qquad
(a\rightarrow0),
\label{app:eq:AdSBH-zeroT-coefficients}
\end{equation}
reproducing the covariant zero-temperature Hamilton--Jacobi functional. We now turn to the scalar sector.

% ------------------------------------------------------------
\subsubsection*{Scalar sector}
% ------------------------------------------------------------

A further check comes from fluctuations of the scalar about $\phi_0$.
Define
\begin{equation}
M(\eta)
\equiv
P_{\phi\phi}(\phi_0,\eta),
\qquad
m_\phi^2
\equiv
V_{\phi\phi}(\phi_0).
\end{equation}
Taking two $\phi$ derivatives of the homogeneous HJ equation gives
\begin{equation}
m_\phi^2
=
M^2
-
\frac34
(P-\eta P_\eta)
\left(
M-\eta M_\eta
\right).
\label{app:eq:M-equation}
\end{equation}
For a scalar sector symmetric about $\phi_0$,
\begin{equation}
F_5(\phi_0,\eta)=0,
\end{equation}
and the scalar-gradient equation becomes
\begin{equation}
\frac34\eta
(P-\eta P_\eta)
F_{4,\eta}
+
\left(
2M-\frac{\lambda_\tau}{2}
\right)F_4
-
\frac12
=
0.
\label{app:eq:F4-AdSBH}
\end{equation}
At the AdS boundary,
\begin{equation}
M(0)=\frac{\delta_*}{\ell},
\qquad
\delta_*(\delta_*-3)=m_\phi^2\ell^2,
\end{equation}
which fixes
\begin{equation}
F_4(0)
=
\frac{\ell}{2(2\delta_*-1)}.
\label{app:eq:F4-boundary}
\end{equation}
This is precisely the scalar-gradient coefficient of the covariant
zero-temperature HJ solution.

The AdS-Schwarzschild case provides a nontrivial check of the two-derivative descent system. The curvature coefficient is obtained exactly, while the Kaluza-Klein and temperature-gradient equations admit smooth asymptotic solutions with the required AdS limits. For a reflection-symmetric scalar sector the mixed coefficient consistently vanishes, and the scalar-gradient equation reproduces the covariant AdS boundary coefficient.
% ------------------------------------------------------------
\subsubsection*{Reconstruction of the AdS-Schwarzschild geometry}
% ------------------------------------------------------------

The homogeneous characteristic equations also allow us to reconstruct
the bulk geometry directly.  For the planar solution we write
\begin{equation}
N=e^{A+\sigma},
\qquad
h_{ij}=e^{2A}\delta_{ij},
\qquad
a_i=0 .
\end{equation}
Using
\begin{equation}
\dot\sigma
=
-\frac12(\lambda_\tau-\lambda_s)
=
-\frac12\eta
=
-\frac{3a}{\ell},\, \frac{da}{du}
=
\frac{3aw}{\ell},\,\frac{d\sigma}{da}
=
-\frac1w .
\end{equation}

Choosing the asymptotic AdS normalization
$\sigma\rightarrow0$ as $a\rightarrow0$, we obtain
\begin{equation}
\sigma
=
-\operatorname{arcsinh}a , \, e^\sigma
=
e^{-\operatorname{arcsinh}a}
=
w-a .
\end{equation}
We can identify the blackening function as
\begin{equation}
f\equiv e^{2\sigma}, \implies f=(w-a)^2
\end{equation}
Since
\begin{equation}
(w-a)(w+a)=1,
\end{equation}
this relation may equivalently be written as
\begin{equation}
2a=\frac{1-f}{\sqrt f}.
\label{app:eq:a-of-f}
\end{equation}

The spatial characteristic equation is
\begin{equation}
\dot A
=
-\frac12\lambda_s
=
-\frac1\ell(w-a)
=
-\frac{\sqrt f}{\ell}.
\label{app:eq:A-AdSBH}
\end{equation}
Defining the usual AdS radial coordinate by
\begin{equation}
r\equiv \ell e^A ,
\end{equation}
we have
\begin{equation}
\frac{\dot r}{r}
=
\dot A
=
-\frac{\sqrt f}{\ell}.
\label{app:eq:rflow-AdSBH}
\end{equation}

The relation between the thermal and spatial flows determines $f(r)$
without any further input.  Indeed,
\begin{equation}
\frac{d\sigma}{dA}
=
\frac{\dot\sigma}{\dot A}
=
\frac{3a}{w-a}.
\end{equation}
Since $f=e^{2\sigma}$,
\begin{equation}
\frac{d\log f}{d\log r}
=
2\frac{d\sigma}{dA}
=
\frac{6a}{w-a}.
\end{equation}
Using \eqref{app:eq:a-of-f} and $w-a=\sqrt f$ gives
\begin{equation}
\frac{r}{f}\frac{df}{dr}
=
3\frac{1-f}{f},
\end{equation}
or
\begin{equation}
r\frac{df}{dr}
=
3(1-f).
\end{equation}
Its asymptotically AdS solution is
\begin{equation}
f(r)
=
1-\frac{r_h^3}{r^3},
\label{app:eq:AdSBH-blackening}
\end{equation}
where the integration constant has been written as $r_h^3$ so that
$f(r_h)=0$.

Finally, \eqref{app:eq:rflow-AdSBH} gives
\begin{equation}
du^2
=
\frac{\ell^2}{r^2f(r)}\,dr^2 .
\end{equation}
The homogeneous metric reconstructed from the Hamilton-Jacobi
characteristics is therefore
\begin{equation}
ds^2
=
\frac{\ell^2}{r^2f(r)}\,dr^2
+
\frac{r^2}{\ell^2}
\left[
f(r)\,d\tau^2
+
dx^2+dy^2
\right],
\qquad
f(r)=1-\frac{r_h^3}{r^3}.
\label{app:eq:AdSBH-reconstructed}
\end{equation}
This is precisely the Euclidean planar AdS$_4$-Schwarzschild
geometry.  Regularity of the Euclidean section at $r=r_h$ fixes
the thermal period to
\begin{equation}
\beta
=
\frac{4\pi\ell^2}{3r_h},
\qquad
T
=
\frac{3r_h}{4\pi\ell^2}.
\label{app:eq:AdSBH-temperature}
\end{equation}

Thus the homogeneous Hamilton-Jacobi solution and its characteristic flow reconstructs the
full planar AdS-Schwarzschild geometry.

\section{Approaching the black hole singularity}
\label{interior}
In the preceding sections, working in Euclidean signature, we established the connection between the Hamiltonian constraint and the dilatation Ward identity at finite temperature and developed the corresponding local Hamilton--Jacobi derivative expansion. Outside the horizon, radial evolution is spacelike and admits the usual holographic interpretation in terms of RG flow. Inside the horizon, however, the radial direction becomes timelike, and the same Hamiltonian framework instead describes time evolution toward the singularity. Thus, to study this regime,  we must work in  Lorentzian signature.
 In this section, we explore the black hole interior, more precisely, the near-singularity behavior, using  Hamilton-Jacobi theory\footnote{See Appendix~\ref{time evolution appendix} details.}.% More precisely, we write the Hamiltonian and momentum constraint near the singularity.% Curiously, the on-shell action is finite near the singularity.

%Using the ADM decomposition, we can write down the projected Einstein equations (see Appendix~\ref{time evolution appendix} details).
 % \begin{eqnarray}
 % \label{eom with matter}
 %     &&\frac{1}{2}\Big(K^{\mu\nu}K_{\mu\nu}-K^2\Big)=\kappa n^{\mu} T^z_{\mu}\\
 %   && -\gamma^{\alpha \mu}\gamma^{\rho\nu}\tilde{\nabla}_{\rho}(K_{\mu\nu}- K \gamma_{ \mu\nu})=\kappa \gamma^{\alpha \mu} T^z_{\mu}=0\\
 %   &&\mathcal{L}_v K_{\mu\nu}= K K_{\mu\nu}- 2 K_{\mu}^{\rho} K_{\rho\nu}-\kappa \gamma^{\alpha}_{\mu}\gamma^{\beta}_{\nu} T_{\alpha\beta}^h+\frac{\kappa}{(d-2)} \gamma_{\mu\nu}[T_{\rho}^z n^{\rho}+T_{\rho\sigma}^h \gamma^{\rho\sigma}].
 %  % && \mathcal{L}_v K_{\mu\nu}= K K_{\mu\nu}- 2 K_{\mu}^{\rho} K_{\rho\nu}+\frac{\kappa}{g}\Big(E_{\mu} E_{\nu}-\frac{h_{\mu\nu}}{d-2} E^{\rho} E_{\rho}\Big)
 % \end{eqnarray}
 % The first equation is the Hamiltonian constraint, while the second one is the momentum constraint. The third equation represents the dynamical equation. The stress tensor on the RHS is from the matter fields.
%We are interested in  the near singularity  behavior. 

%In the near-singularity regime,  the radial derivatives (time-like) dominate over  the spatial derivatives and the equations of motion instead of being  PDEs become ODEs. 
In the near-singularity regime, the time-like radial derivatives dominate over the spatial ones, and the equations of motion reduce from partial differential equations to ordinary differential equations.
Belinski, Khalatnikov and Lifshitz (BKL) showed that in Einstein gravity,  the approach to the singularity is generically oscillatory and chaotic \cite{Belinsky:1970ew,LIFSHITZ1992659}. For example, in AdS, it was recently shown that including three gauge fields leads to mixmaster dynamics close to the singularity \cite{DeClerck:2023fax}. %\cite{PhysRev.186.1319,PhysRevLett.22.1071}.  
However, in theories involving only scalar fields, the approach to the singularity is non-chaotic and exhibits Kasner-like behavior. The holographic implications of such scenarios have been recently discussed in \cite{Frenkel:2020ysx,Caceres:2022hei,Caceres:2023zft,Caceres:2023zhl,Caceres:2024edr,Caceres:2022smh,DeClerck:2023fax}.

% \subsection{Solutions of EOM and BKL}
%We solve the equation of motion with just a single scalar field.  The dynamics with 3 gauge fields are much more interesting and show mix-master dynamics\cite{DeClerck:2023fax,PhysRev.186.1319,PhysRevLett.22.1071}. 

%In \cite{DeClerck:2023fax,PhysRev.186.1319,PhysRevLett.22.1071}, the authors showed that including 3 gauge fields the dynamic is chaotic and exhibits mixmaster behavior. 
Let us consider Einstein gravity with a minimally coupled massive scalar and use a Kasner-like ansatz,
%To make a connection with BKL dynamics, the authors \cite{Oling:2024vmq} took the following ansatz (inspire by Kasner metric) for the $n^{\mu}$ and spatial metric $\gamma_{\mu\nu}$\footnote{We are inside the horizon hence the radial coordinate $z$ is time like and ``time'' coordinate is spacelike.}.
\begin{eqnarray}
 ds^2= - e^{\alpha (z)} dz^2 + \gamma_{\mu\nu}dx^{\mu}dx^{\nu}%= e^{2 \beta_x(z)}dx^2+e^{2 \beta_y(z)}dy^2+e^{2 \beta_t(z)}dt^2
\end{eqnarray}
where the spatial metric $\gamma_{\mu\nu}$ is\footnote{Inside the black hole  the $t$ coordinate is spacelike.},
\begin{eqnarray}
\label{hypersurface ansatz}
  %  n^{\mu} \partial_{\mu}= - e^{\alpha (z)}\partial_z, \quad 
  \gamma_{\mu\nu}dx^{\mu}dx^{\nu}= e^{2 \beta_x(z)}dx^2+e^{2 \beta_y(z)}dy^2+e^{2 \beta_t(z)}dt^2
\end{eqnarray}
With this ansatz, we can solve Einstein's equations and obtain the evolution of the exponents. %These exponents shows interesting behaviors which is discussed in those references\cite{Oling:2024vmq}. 
Our first goal is to understand how the on-shell action depends on these exponents. To this effect, we turn to Hamilton-Jacobi theory. But before doing so, let us review the case of an AdS-Schwarzschild black hole as a way of establishing the notation.

 The metric of AdS-Schwarzschild is,
   \begin{eqnarray}
       ds^2&&= \frac{L^2}{z^2}\Big(-(1-z^3/z_0^3)dt^2+\frac{dz^2}{1-z^3/z_0^3}+dx^2+dy^2\Big)\nonumber,
      \end{eqnarray}
 where $z_0= 3 \beta/4 \pi$ and the boundary is at $z\rightarrow 0$. 
Near the singularity at $z \rightarrow \infty $,
      \begin{eqnarray}
          ds^2  \sim L^2\left(-d\tau^2+(\frac{2}{3})^{2/3}\frac{1}{z_0^2} \tau^{-2/3} dt^2+\frac{\tau^{4/3}}{(\frac{2}{3})^{4/3}z_0^2}[dx^2+dy^2]\right).
      \end{eqnarray}
Where the radial coordinate $\tau=\tau(z)$ is now time-like. This is a Kasner metric with $p_t=-\frac{1}{3}$ and $p_x=p_y= 2/3$. We can absorb numerical factors by further coordinate redefinitions.  %redefine other coordinates to absorb the factor of $2/3$,
On a constant $\tau$ hypersurface the metric is\footnote{We can write this  metric in the form of \eqref{hypersurface ansatz} by identifying $
 \tau^{-2/3}= e^{2 \beta_t},\,    \tau^{4/3}= e^{2 \beta_x}, \, \beta_x=\beta_y$.
},
 \begin{eqnarray}
      \gamma_{\mu\nu}dx^{\mu} dx^{\nu} \sim \tau^{-2/3} dt^2+\tau^{4/3}[dx^2+dy^2].
  \end{eqnarray}
In general, a vacuum solution of Kasner form has exponents that satisfy, 
   \begin{eqnarray}
   p_t+p_x+p_y=1, \quad %\sum_i p_i^2=
   p_t^2+p_x^2+p_y^2=1
\end{eqnarray}
In the presence of a scalar, the Kasner relations are modified, and we have, 
$$p_t+p_x+p_y=1,\qquad  p_\phi^2+p_t^2+p_x^2+p_y^2+=1,$$
where $p_\phi$ is related to the near-singularity behavior of the scalar field, $\phi\sim-\sqrt{2}p_\phi\log \tau.$

%We can also write the above metric in the form of \eqref{hypersurface ansatz} by identifying 
%\begin{eqnarray}
% \tau^{-2/3}= e^{2 \beta_t},\quad    \tau^{4/3}= e^{2 \beta_x}, \quad  \beta_x=\beta_y.
%\end{eqnarray}
% This gives $\bar{\beta}_a^{(0)}=0$ and $\bar{v}_1=\frac{\sqrt{\frac{2}{3}} \log (\tau )}{\tau }, \,\, \bar{v}_2=-\frac{\sqrt{\frac{2}{3}} \log (\tau )}{\tau }, \,\,  \bar{v}_3=0$. This also gives $\beta_a$ as
% \begin{eqnarray}
%     \beta_a=\left\{\frac{2 \log (\tau )}{3},\frac{2 \log (\tau
%    )}{3},-\frac{\log (\tau )}{3}\right\}
% \end{eqnarray}
% Now we can define $\tau^{4/3}=1/z^2$ and write the metric on the hypersurface as
% \begin{eqnarray}
%     h_{ij}dx^i dx^j= \frac{1}{z^2} \Big(z^3 dt^2+dx^2+dy^2\Big)
% \end{eqnarray}
% This is the form of the metric near the singularity $z>>1$. The Carrollian ansatz (scaling argument) will break at some finite $\tau=\tau_0$ and we won't be able to continue this process to the horizon. In the next section, we will see how we can extend this framework to the horizon.\\

\subsection{Hamilton Jacobi analysis in the interior}
Consider the ansatz,
%\begin{eqnarray}
%\label{kasner}
%    n^{\mu} \partial_{\mu}=  -e^{\alpha(z)}\partial_z, \quad \gamma_{\mu\nu}dx^{\mu}dx^{\nu}= e^{2 \beta_x(z)}dx^2+e^{2 \beta_y(z)}dy^2+e^{2 \beta_t(z)}dt^2, \quad \phi= \phi(z)
%\end{eqnarray}
\begin{align}
\label{kasner}
   &n^{\mu} \partial_{\mu}=  -e^{\alpha(z)}\partial_z, \quad \gamma_{\mu\nu}dx^{\mu}dx^{\nu}= e^{2 \beta_x(z)}dx^2+e^{2 \beta_y(z)}dy^2+e^{2 \beta_t(z)}dt^2, \quad \phi= \phi(z)\nonumber\\
  & 
\end{align}
%With the above ansatz, one can write the Hamiltonian and momentum constraint in terms of individual metric components. Still, it is more feasible 
Let us first write the on-shell action in terms of metric variables and identify the dynamical variables. The action is
\begin{eqnarray}
    S=\int d^4 x \sqrt{-g} R+2 \int d^3 x\, \sqrt{\gamma} K+S_m,
\end{eqnarray}
where 
\begin{eqnarray}
 S_{matter}=\int \sqrt{-g}\, d^4 x\,\ \Big(-\frac{1}{2} g^{ab} \partial_{a}\phi \,\partial_{b}\phi+ V(\phi)\Big),   \end{eqnarray}
 and we have set $\frac{1}{16 \pi G}=1$.
The signs of various terms are consistent with the Lorentzian signature of the metric and the potential $V(\phi)\sim \frac{6}{\ell^2}-\frac{1}{2} m^2 \phi^2 +\cdots $ contains the cosmological constant.
Using the ADM decomposition of Appendix~\ref{time evolution appendix}, we write the Einstein-Hilbert term as,
\begin{eqnarray}
    S=\int d^4 x\,  \sqrt{\gamma}\,N [{}^3R+K_{\mu\nu}K^{\mu\nu}-K^2]+S_{matter}.
\end{eqnarray}
In terms of metric components, using \eqref{kasner}, this  yields the on-shell action,
% \footnote{One can verify that starting from the Einstein-Hilbert term plus the Gibbons-Hawking-York boundary term 
% \begin{eqnarray}
%     S_{bulk}&&=\int dz\, V_3\, e^{-\alpha_z(z)+\beta_t(z)+\beta_x(z)+\beta_y(z)} \Big(\frac{1}{2}\phi'(z)^2- V e^{2 \alpha_z(z)}-2
%    \alpha_z'(z) \left(\beta_t'(z)+\beta_x'(z)+\beta_y'(z)\right)\nonumber\\
%   && +2 \left(\beta_t''(z)+\beta_t'(z) \left(\beta_x'(z)+\beta_y'(z)\right)+\beta_t'(z)^2+\beta_x''(z)+\beta_x'(z) \beta_y'(z)+\beta_x'(z)^2+\beta_y''(z)+\beta_y'(z)^2\right)\Big)\nonumber\\
% \end{eqnarray}
% \begin{eqnarray}
%   S_{bdry}=  2 V_3\, e^{-\alpha_z(z)+\beta_t(z)+\beta_x(z)+\beta_y(z)} \left(\beta_t'(z)+\beta_x'(z)+\beta_y'(z)\right)\Big|_{z=z_0}.
% \end{eqnarray}
% And upon integration by parts, one can arrive at the above result.}
\begin{eqnarray}
    &&S= -2\int d^4 x\,  \,\sqrt{\gamma} Ne^{2 \alpha(z)}\left(\beta_t'(z) \left(\beta_x'(z)+\beta_y'(z)\right)+\beta_x'(z) \beta_y'(z)\right)+S_{matter}\nonumber\\
  \end{eqnarray}
 where   $\sqrt{\gamma}N=e^{-\alpha(z)+\beta_t(z)+\beta_x(z)+\beta_y(z)}.$ % The matter action  to be added to teh Einstein-Hilbert action is,
  %\begin{eqnarray}
   %   S_m= \int d^4 x \sqrt{\gamma} N [\frac{1}{2} g^{zz}\partial_z \phi \partial_z \phi +V(\phi)- \frac{1}{2}\gamma^{\mu\nu} \partial_{\mu} \phi\partial_{\nu}  \phi].
 % \end{eqnarray}
Adding the matter contribution, we obtain, 
\begin{eqnarray}
\label{lagrangian}
    S= \int d^4 x\,  \sqrt{\gamma} N\, \Bigg[e^{2 \alpha(z)} \left(- \bar{\beta}_t'(z)^2 +\bar{\beta}_x'(z)^2+\bar{\beta}_y'(z)^2\right)+\left(\frac{1}{2}g^{zz}\partial_z \phi \partial_z \phi  +V(\phi)\right)\Bigg]\nonumber\\
\end{eqnarray}

Note that in \eqref{lagrangian} have redefined the $\beta$'s as follows 
\begin{eqnarray}
   && \beta_t=\frac{1}{\sqrt{6}}(\bar{\beta}_t-\bar{\beta}_x-\sqrt{3}\bar{\beta}_y)\nonumber\\
   && \beta_x=\frac{1}{\sqrt{6}}(\bar{\beta}_t-\bar{\beta}_x+\sqrt{3}\bar{\beta}_y)\nonumber\\
   && \beta_y=\frac{1}{\sqrt{6}}(\bar{\beta}_t+2\bar{\beta}_x).
\end{eqnarray}

With these redefinitions,  $\sqrt{\gamma}=\frac{\sqrt{3}}{\sqrt{2}}\bar{\beta_t}(z)$. And we also assume that the matter field only depends on the radial coordinate as $\phi(z)$. Now, the conjugate momenta for the above variables are defined as, %\footnote{One could have further redefined $\beta$'s to absorb the factor of $e^{2\alpha(z)}$. But we keep it here explicitly. We also put $N \sqrt{\gamma}$ in the Lagrangian $L$.}
\begin{eqnarray}
    \pi_{\bar{\beta}_t}= \frac{\partial L}{\partial \, \bar{\beta}_t'}=-2\bar{\beta}_t'e^{ \alpha(z)}\sqrt{\gamma}, \quad  \pi_{\bar{\beta}_x}= \frac{\partial L}{\partial \, \bar{\beta}_x'}\sqrt{\gamma}=2\bar{\beta}_x'e^{ \alpha(z)},\quad  \pi_{\phi}= \frac{\partial L}{\partial \, \phi'}=\phi'e^{ \alpha(z)}\sqrt{\gamma}\nonumber.\\
\end{eqnarray}
Thus, the Hamiltonian constraint for the ansatz \eqref{kasner},
\begin{align}
 \mathcal{H}&= \mathcal{H}_{grav} + \mathcal{H}_{matter}=0\nonumber\\
 &=K^2- K_{ij}K^{ij}+ \mathcal{H}_{matter}=0
 \end{align}
becomes,\footnote{In general, the Hamiltonian constraint is given by the Legendre transform of the Lagrangian, \\
%\begin{eqnarray*}
 $   \mathcal{H}= \pi^{ij} \dot{\gamma_{ij}}-L
  =-\sqrt{\gamma}\Big[N\, C_0- 2 \beta^i C_i\Big] $
%\end{eqnarray*}
with $C_0:=R +K^2- K_{ij}K^{ij}$ and $C_i:=D_j K^{j}_i- D_i K$
.}% in terms of these conjugate variables as
\begin{eqnarray}
\label{kasnerhamilcons}
 \frac{e^{- \alpha(z)}}{\sqrt{\gamma}}  \Big[  -\frac{1}{4}\pi_{\bar{\beta}_t}^2+\frac{1}{4}\pi_{\bar{\beta}_x}^2+\frac{1}{4}\pi_{\bar{\beta}_y}^2+\frac{1}{2} \pi_{\phi}^2-\gamma V(\phi)\Big]=0 
\end{eqnarray}
%We have four dynamical variables $\bar{\beta}_t,\bar{\beta}_x,\bar{\beta}_y,\phi'$. 
Next, we use Hamilton-Jacobi theory to write conjugate momenta as the variation of the on-shell action with respect to dynamical fields $\bar{\beta}_t,\bar{\beta}_x,\bar{\beta}_y$ and $\phi'$, 
\begin{eqnarray}
    \pi_{\bar{\beta}_t}=\frac{\delta S}{\delta\, \bar{\beta}_t },\quad \pi_{\bar{\beta}_x}=\frac{\delta S}{\delta\, \bar{\beta}_x },\quad \pi_{\bar{\beta}_y}=\frac{\delta S}{\delta\, \bar{\beta}_y },\quad \pi_{\phi}=\frac{\delta S}{\delta \phi}
\end{eqnarray}
Hence, the Hamiltonian constraint becomes
\begin{align}
\label{bklhamilconstra}
\boxed{
\frac{e^{- \alpha(z)}}{\sqrt{\gamma}} \Bigg(      -\frac{1}{4}\Big(\frac{\delta S}{\delta\, \bar{\beta}_t }\Big)^2+\frac{1}{4}\Big(\frac{\delta S}{\delta\, \bar{\beta}_x }\Big)^2+\frac{1}{4}\Big(\frac{\delta S}{\delta\, \bar{\beta}_y }\Big)^2+\frac{1}{2} \Big(\frac{\delta S}{\delta \phi}\Big)^2-V(\phi) \gamma \Bigg)=0 }.
\end{align}

One could easily rewrite this equation in terms of the Kasner exponents instead of the $\bar\beta_i$'s. Two remarks are in order. 
\begin{enumerate}
    \item Clearly, given a boundary deformation, $I_\mathcal{O}= \int d^3x \phi_0 \mathcal{O},$ the boundary parameter $\phi_0$ and the corresponding Kasner exponent $p_\phi$ are related via the equations of motion.  
In a background with boundary planar symmetry, knowing one exponent determines all the other ones, since in that case we have $p_t+2 p_x =1$ and $p_\phi^2 + p_t^2 + 2p_x^2=1.$ Thus, we can think of the Kasner exponents as functions of the boundary data and of the near-singularity Hamiltonian constraint \eqref{bklhamilconstra} as implicitly related to the boundary deformation governed by $\phi_0$\footnote{More precisely, the dimensionless parameter $\frac{\phi_0}{T^{d-\Delta}}$\cite{Caceres:2021fuw,Frenkel:2020ysx}.}.
\item If we quantize the Hamiltonian constraint in the interior by promoting the momenta to operators, we obtain the Wheeler-deWitt equation. The solution of the WdW equation in a semi-classical regime is just $\Psi=e^{i\,S}.$ Thus, if we can solve \eqref{bklhamilconstra} for $S[\bar{\beta}_t ,\bar{\beta}_x,\bar{\beta}_y]$
 we have solved the WdW equation. It can be shown  that if there is no matter, just a potential 
$V=\frac{6}{\ell^2}$, there exists an exact solution of \eqref{bklhamilconstra}. The solution is, 
%$S[\bar{\beta}_t ,\bar{\beta}_x,\bar{\beta}_y],$ can be written as
\begin{eqnarray}
\label{Sbetabar}
  S[\bar{\beta}_t ,\bar{\beta}_x,\bar{\beta}_y]=4 e^{\sqrt{\frac{3}{2}} \bar{\beta}_t} \sinh \left(c_1
   -\frac{\sqrt{3} \bar{\beta}_x+3 \bar{\beta}_y}{2
   \sqrt{2}}\right),
\end{eqnarray}
where $c_1$ is a constant of integration which can be fixed in terms of the energy (ADM mass) of the system. The existence of this solution allowed \cite{Hartnoll:2022snh} to formulate the interior of a black hole as a superposition of WdW states. In our case, when the scalar field $\phi$ is non-zero, we could similarly try to find the on-shell action as a functional of these exponents. We leave the study of this scenario and its implications for future work.\\
\end{enumerate}

    \section{Discussion}
    \label{discussions}
   The seminal work of de Boer, Verlinde, and Verlinde  \cite{deBoer:1999tgo} showed that the Callan-Symanzik equation for vacuum CFTs can be derived from a holographic RG flow framework. In the present work, we focused on thermal CFTs and studied the Hamiltonian constraint both in the exterior and interior of a black hole. In the exterior, we obtained the Ward identity for broken dilatations directly from the Hamiltonian constraint. This extends the zero-temperature Hamilton-Jacobi construction of \cite{deBoer:1999tgo} to a finite-temperature background. The key new ingredient is the conserved thermal charge \(D\), which captures the anisotropy between the Euclidean time circle and the spatial directions. A naive replacement \(U(\phi)\rightarrow U(\phi,T)\) would miss this additional data. Using parameter \(D\), we constructed the stationary Hamilton-Jacobi functional upto two derivatives. Its coefficients obey a closed set of radial descent equations, while the momentum constraint fixes how the local thermal branch can vary in space. Separating the local terms from the non-local functional \(\Gamma_D\) then gives a finite-\(D\) functional Callan-Symanzik equation. In the ultraviolet, where the thermal anisotropy disappears, the construction smoothly reduces to the usual covariant zero-temperature Hamilton-Jacobi expansion.
Our final contribution is related to the interior of black holes. Recall that the near-singularity dynamics can be understood in terms of BKL exponents, which generically show mixmaster behavior. Using Hamilton-Jacobi theory, we found a differential equation for the on-shell action as a functional of these exponents. Subsequently, using the relationship between the exponents and boundary data, this equation can be interpreted as relating near-singularity dynamics and evolution in RG space. 
   
   Our work points to several directions that deserve further study.    
   %pursue.
    \begin{itemize}
        \item  The near-singularity dynamics is known to receive corrections from higher derivative terms like $R^2,R^3$, etc. \cite{Bueno:2024fzg, Bueno:2024qhh}. Furthermore, the inclusion of matter distinctly alters this behavior \cite{Caceres:2024edr}. Thus, it is natural to ask if the equation we derived here can shed some light on the physics of eons and epochs, in the presence of matter, from the boundary perspective. 
 
        \item Einstein's equations, which are generally  PDEs, become ODEs near the black hole singularity. This is due to the ultralocal behavior of the coordinates (the radial derivative $z$ matters more than the spatial derivative). This regime, dubbed Carrollian gravity,  is a $c\rightarrow 0$ limit of general relativity \cite{Oling:2024vmq,Musaeus:2023oyp,Hansen:2021fxi}. 
        The on-shell action that one gets from Carrollian expansion is precisely the same as that of the ADM formalism (at the leading order). Hence, one can try to pursue the quantization of Carrollian gravity, which is much simpler than Einstein GR. 
        One significant open question is understanding the mixmaster dynamics in terms of (boundary) coupling space dynamics.
        Solving this in generality is clearly a challenging task, but borrowing intuition from the previous point,  it might be possible to find a Carrollian sub-sector in holographic  QFTs and its RG flows at finite temperature.

        \item  Using the ADM decomposition, we have found the Hamiltonian constraint in the interior of a black hole. The quantized form of this equation is the Wheeler-deWitt (WdW) equation. In \cite{McGough:2016lol}, the authors formulated the WdW equation in terms of a  $T\bar{T}$ deformation of the boundary theory. It would be extremely interesting to generalize their setup to understand the $T\bar{T}$-like setup to probe the interior of the black hole. Some aspects of this setup have been studied in  \cite{AliAhmad:2025kki}, but there is much more to be explored. 
        
    \end{itemize}

    \section*{Acknowledgements}
     We thank G.~Oling, H.~Verlinde, and G.~Cuomo for useful discussions. E.C. thanks the Instituto de Física Teórica, Madrid, Spain, for hospitality and the participants of the ``New Insights in Black Hole Physics from Holography" workshop for exciting discussions. The work of E.C. and H.K. is supported by NSF grant PHY-2210562 and CNS Spark Grant 2025-2029.

% ============================================================

\appendix
\section{Technical details of the thermal Hamilton-Jacobi construction}
\label{app:thermal-HJ}

This appendix gives the intermediate steps underlying the thermal
Hamilton-Jacobi equations used in the main text.  We derive the
homogeneous fixed-$D$ equation, relate it to the $(W,f)$ formulation,
and summarize the canonical ingredients required at two derivatives.

% ============================================================
\subsection{Homogeneous fixed D Hamilton-Jacobi equation}
\label{app:zeroth-HJ}
% ============================================================

For the homogeneous ansatz, define
\begin{equation}
\rho\equiv 3A+\sigma .
\end{equation}
The reduced action is
\begin{equation}
S_{\rm red}
=
\beta V_2
\int du\,e^\rho
\left[
-\frac23\dot\rho^{\,2}
+\frac23\dot\sigma^{\,2}
+\frac12\dot\phi^2
+V(\phi)
\right].
\label{app:eq:reduced-action}
\end{equation}
The corresponding canonical momenta are
\begin{equation}
p_\rho
=
-\frac43\beta V_2e^\rho\dot\rho,
\qquad
p_\sigma
=
\frac43\beta V_2e^\rho\dot\sigma,
\qquad
p_\phi
=
\beta V_2e^\rho\dot\phi .
\label{app:eq:canonical-momenta}
\end{equation}

Because $\sigma$ is cyclic, its momentum is conserved.  We parameterize
the conserved value by
\begin{equation}
p_\sigma=-\frac23\beta V_2D ,
\end{equation}
so that
\begin{equation}
\dot\sigma
=
-\frac12D e^{-\rho}.
\label{app:eq:sigma-dot-D}
\end{equation}
Introducing
\begin{equation}
\eta\equiv D e^{-\rho},
\end{equation}
this becomes simply
\begin{equation}
\dot\sigma=-\frac12\eta .
\end{equation}

The reduced Hamiltonian is
\begin{equation}
H
=
\frac{1}{\beta V_2}
\left[
-\frac38e^{-\rho}p_\rho^2
+\frac38e^{-\rho}p_\sigma^2
+\frac12e^{-\rho}p_\phi^2
\right]
-\beta V_2e^\rho V(\phi).
\label{app:eq:reduced-Hamiltonian}
\end{equation}
Writing
\begin{equation}
S=\beta V_2\,\mathcal S ,
\end{equation}
the Hamilton-Jacobi constraint $H=0$ becomes
\begin{equation}
-\frac38e^{-\rho}
(\partial_\rho\mathcal S)^2
+
\frac38e^{-\rho}
(\partial_\sigma\mathcal S)^2
+
\frac12e^{-\rho}
(\partial_\phi\mathcal S)^2
-
e^\rho V
=
0.
\label{app:eq:reduced-HJ}
\end{equation}

On a fixed-$D$ branch, take
\begin{equation}
\mathcal S_D^{(0)}
=
-\frac23D\sigma
+
e^\rho P(\phi,\eta),
\qquad
\eta=D e^{-\rho}.
\label{app:eq:fixedD-characteristic}
\end{equation}
Its derivatives are
\begin{equation}
\partial_\rho\mathcal S_D^{(0)}
=
e^\rho(P-\eta P_\eta),
\qquad
\partial_\sigma\mathcal S_D^{(0)}
=
-\frac23D,
\qquad
\partial_\phi\mathcal S_D^{(0)}
=
e^\rho P_\phi .
\end{equation}
Substitution into \eqref{app:eq:reduced-HJ} gives
\begin{equation}
V(\phi)
=
\frac12P_\phi^2
-\frac38(P-\eta P_\eta)^2
+\frac16\eta^2 .
\label{app:eq:thermal-HJ-homogeneous}
\end{equation}

The corresponding characteristic equations are
\begin{equation}
\dot\rho
=
-\frac34(P-\eta P_\eta),
\qquad
\dot\sigma
=
-\frac12\eta,
\qquad
\dot\phi
=
P_\phi.
\label{app:eq:homogeneous-characteristics}
\end{equation}
Using $\rho=3A+\sigma$, these reproduce the homogeneous radial flow
used in the main text.

% ============================================================
\subsection*{Relation to the $(W,f)$ first-order system}
\label{app:Wf-map}
% ============================================================

The same homogeneous dynamics may be written in terms of the
blackening function
\begin{equation}
f=e^{2\sigma}
\end{equation}
and a first-order function $W(\phi)$.  Along a homogeneous branch,
\begin{equation}
\dot A=-\frac14e^\sigma W,
\qquad
\dot\phi=e^\sigma W_\phi .
\label{app:eq:Wf-flow}
\end{equation}
Matching these equations to
\eqref{app:eq:homogeneous-characteristics} gives
\begin{equation}
e^\sigma W
=
P-\eta P_\eta-\frac23\eta,
\qquad
e^\sigma W_\phi=P_\phi .
\label{app:eq:PW-map}
\end{equation}

Since
\begin{equation}
\sigma_\phi=\frac{f_\phi}{2f},
\end{equation}
we have
\begin{equation}
\dot\sigma
=
\sigma_\phi\dot\phi
=
\frac12e^{-\sigma}f_\phi W_\phi .
\end{equation}
Comparing with $\dot\sigma=-\eta/2$ yields
\begin{equation}
\eta
=
-e^{-\sigma}f_\phi W_\phi .
\label{app:eq:eta-Wf}
\end{equation}
Together with $\eta=D e^{-3A-\sigma}$, this implies
\begin{equation}
f_\phi W_\phi
=
-D e^{-3A}.
\label{app:eq:blackening-D}
\end{equation}

Using
\begin{equation}
\frac{dA}{d\phi}
=
-\frac{W}{4W_\phi},
\end{equation}
differentiation of \eqref{app:eq:blackening-D} gives
\begin{equation}
f_{\phi\phi}W_\phi
+
f_\phi W_{\phi\phi}
-
\frac34Wf_\phi
=
0,
\label{app:eq:blackening-scalar}
\end{equation}
which is the blackening equation in scalar-field variables.

Finally, substituting
\begin{equation}
P_\phi=e^\sigma W_\phi,
\qquad
P-\eta P_\eta=e^\sigma W+\frac23\eta,
\qquad
\eta=-e^{-\sigma}f_\phi W_\phi
\end{equation}
into \eqref{app:eq:thermal-HJ-homogeneous} gives
\begin{equation}
\left(
\frac12W_\phi^2-\frac38W^2
\right)f
+
\frac12WW_\phi f_\phi
-
V(\phi)
=
0.
\label{app:eq:Wf-constraint}
\end{equation}
Thus the fixed-$D$ function $P(\phi,\eta)$ and the pair $(W,f)$ are
two equivalent descriptions of the same homogeneous radial dynamics.

% ============================================================
\subsection*{Two-derivative canonical descent}
\label{app:two-derivative-details}
% ============================================================

We now summarize the intermediate canonical steps leading to the
five two-derivative descent equations of the main text.

% ------------------------------------------------------------
\subsubsection*{Canonical responses}
% ------------------------------------------------------------

Let
\begin{equation}
\mathcal E,\qquad
\mathcal T^i{}_j,\qquad
\mathcal J_i,\qquad
\mathcal O
\end{equation}
denote the lapse, spatial-metric, Kaluza--Klein, and scalar responses
of the two-derivative functional.  Their canonical momenta are
\begin{equation}
\Pi^\tau{}_{\tau(2)}
=
\frac12\mathcal E,
\qquad
\Pi^i{}_{j(2)}
=
\frac12\mathcal T^i{}_j,
\qquad
\Pi^\tau{}_{i(2)}
=
\frac{1}{2N^2}\mathcal J_i,
\qquad
\Pi_{\phi(2)}
=
\mathcal O .
\label{app:eq:response-momenta}
\end{equation}
Using
\begin{equation}
K^\mu{}_\nu
=
\Pi^\mu{}_\nu
-\frac12\Pi\,\delta^\mu{}_\nu ,
\end{equation}
these responses determine the second-order corrections to the radial
characteristics.

% ------------------------------------------------------------
\subsubsection*{Hamiltonian constraint at two derivatives}
% ------------------------------------------------------------

At second derivative order the Hamiltonian constraint reduces to
\begin{equation}
0=
\int d^2x\sqrt h\,N
\left[
-\frac12\lambda_\tau\mathcal E
-\frac12\lambda_s\mathcal T
+
P_\phi\mathcal O
+
R[h]
-\frac14N^2\mathcal F^2
-\frac12(\nabla\phi)^2
\right].
\label{app:eq:HJ2-response}
\end{equation}
Here
\begin{equation}
\mathcal T\equiv\mathcal T^i{}_i .
\end{equation}

The terms involving spatial derivatives of
$\lambda_\tau$, $\lambda_s$, and $P_\phi$ arise from divergence terms
in the functional responses.  After the functional variation is taken,
integration by parts gives
\begin{align}
-\frac12\lambda_\tau\mathcal E\big|_{\rm div}
&\longrightarrow
-F_3\nabla^i\log N\,\nabla_i\lambda_\tau
-\frac12F_5\nabla^i\phi\,\nabla_i\lambda_\tau,
\label{app:eq:IBP-E}
\\
-\frac12\lambda_s\mathcal T\big|_{\rm div}
&\longrightarrow
-\left[
F_1\nabla^i\log N+\nabla^iF_1
\right]
\nabla_i\lambda_s,
\label{app:eq:IBP-T}
\\
P_\phi\mathcal O\big|_{\rm div}
&\longrightarrow
\left[
2F_4\nabla^i\phi
+
F_5\nabla^i\log N
\right]
\nabla_iP_\phi .
\label{app:eq:IBP-O}
\end{align}
These terms are the origin of the hydrostatic derivatives appearing
in the final $F_3$, $F_4$, and $F_5$ equations.

% ------------------------------------------------------------
\subsubsection*{Momentum constraint and hydrostatic reduction}
% ------------------------------------------------------------

The leading spatial momentum constraint is
\begin{equation}
\eta
\left[
-\frac12P_{\phi\eta}\partial_i\phi
+
\left(
\frac13-\frac12Q_\eta
\right)\partial_i\sigma
-
\frac32Q_\eta\partial_iA
+
\left(
\frac16+\frac12Q_\eta
\right)\partial_i\log D
\right]
=0,
\label{app:eq:general-momentum}
\end{equation}
where
\begin{equation}
Q\equiv P-\eta P_\eta.
\end{equation}
Thus $D$ is conserved along the radial direction but need not be
spatially constant.

On the reference slice
\begin{equation}
\nabla_iA=0,
\end{equation}
the momentum constraint gives
\begin{equation}
\nabla_i\log\eta
=
\frac{3}
{1+3Q_\eta}
\left(
P_{\phi\eta}\nabla_i\phi-\nabla_i\sigma
\right).
\label{app:eq:hydrostatic-eta}
\end{equation}
Consequently, for any coefficient function $G(\phi,\eta)$,
\begin{equation}
\nabla_i G
=
\left(\mathscr D^{\rm h}_\phi G\right)\nabla_i\phi
+
\left(\mathscr D^{\rm h}_\sigma G\right)\nabla_i\sigma ,
\label{app:eq:hydrostatic-chain-rule}
\end{equation}
with
\begin{equation}
\mathscr D^{\rm h}_\phi
=
\partial_\phi
+
\frac{3\eta P_{\phi\eta}}
{1+3Q_\eta}\partial_\eta,
\qquad
\mathscr D^{\rm h}_\sigma
=
-\frac{3\eta}
{1+3Q_\eta}\partial_\eta .
\label{app:eq:hydrostatic-operators}
\end{equation}
These relations are imposed only after the functional variation has
been performed.  Matching the independent two-derivative structures
then gives the five descent equations quoted in the main text.

% ------------------------------------------------------------
\subsection{Zero-temperature consistency of the descent equations}
\label{app:zeroT-five-equations}
% ------------------------------------------------------------

The finite-temperature descent system provides a nontrivial check when
$\eta\rightarrow0$.  The five equations reduce to
\begin{align}
0={}&
U_\phi F_{1,\phi}
-\frac14UF_1
+1,
\label{app:eq:F1-zero}
\\
0={}&
U_\phi F_{2,\phi}
-\frac14UF_2
-\frac14,
\\
0={}&
U_\phi F_{3,\phi}
-\frac14UF_3,
\\
0={}&
U_\phi F_{4,\phi}
+
\left(
2U_{\phi\phi}-\frac14U
\right)F_4
-\frac14U_\phi F_5
-\frac12U_\phi F_{1,\phi}
-\frac12,
\\
0={}&
U_\phi F_{5,\phi}
+
\left(
U_{\phi\phi}-\frac14U
\right)F_5
-\frac12U_\phi F_1-\frac{1}{2} U_{\phi} F_3
\label{app:eq:F5-zero}
\end{align}

Three-dimensional covariance at zero temperature requires
\begin{equation}
F_2=-\frac14F_1,
\qquad
F_3=0,
\qquad
F_5=2F_{1,\phi},
\qquad
F_4=
\frac{U}{2U_\phi}F_{1,\phi}.
\label{app:eq:zeroT-relations}
\end{equation}

To see that these relations solve the entire system, define
\begin{equation}
\mathcal C_1
\equiv
U_\phi F_{1,\phi}
-\frac14UF_1
+1.
\end{equation}
The remaining four equations then reduce to
\begin{equation}
\mathcal C_2=-\frac14\mathcal C_1,
\qquad
\mathcal C_3=0,
\end{equation}
and
\begin{equation}
\mathcal C_4
=
\frac{U}{2U_\phi}\partial_\phi\mathcal C_1
-\frac12\mathcal C_1,
\qquad
\mathcal C_5
=
2\partial_\phi\mathcal C_1.
\label{app:eq:zeroT-collapse}
\end{equation}
Hence all five equations collapse to the single covariant curvature
descent equation
\begin{equation}
\mathcal C_1=0.
\end{equation}
At zero temperature, our second-order system reduces to the standard
de Boer-Verlinde-Verlinde \cite{deBoer:1999tgo,deBoer:2000cz} Hamilton-Jacobi structure for a
three-dimensional boundary theory. 
Fukuma, Matsuura and Sakai generalize the
this HJ method to arbitrary $d$ \cite{fukuma2000weyl}. Their local functional
through two derivatives is written as
\begin{equation}
S_{\rm loc}
=
\int d^dx\,\sqrt{G}
\left[
W(\phi)
-\Phi(\phi)R[G]
+\frac12 M(\phi)(\nabla\phi)^2
\right].
\end{equation}
For a single canonically normalized scalar, the corresponding
two-derivative descent equations include
\begin{equation}
-1
=
\frac{d-2}{2(d-1)}W\Phi
-
W_\phi\Phi_\phi,
\qquad
W\Phi_\phi+W_\phi M=0.
\end{equation}
Specializing to $d=3$ and identifying our conventions as
\begin{equation}
W=U,
\qquad
\Phi=-F_1,
\qquad
M=2F_4,
\end{equation}
the first relation becomes
\begin{equation}
U_\phi F_{1,\phi}
-\frac14UF_1
+1=0,
\end{equation}
which is precisely our zero-temperature curvature descent equation.
The second relation gives
\begin{equation}
2U_\phi F_4-UF_{1,\phi}=0,
\end{equation}
or equivalently
\begin{equation}
F_4
=
\frac{U}{2U_\phi}F_{1,\phi}.
\end{equation}
Thus the zero-temperature limit of our finite-temperature
Hamilton-Jacobi system reproduces the standard three-dimensional
covariant HJ descent relations. 

% In the covariant formulation the local
% two-derivative functional contains only
% \begin{equation}
% S_{\rm loc}^{(2)}
% =
% \int d^3x\,\sqrt{\gamma}
% \left[
% \Phi(\phi)R[\gamma]
% +\frac12 M(\phi)(D\phi)^2
% \right],
% \end{equation}
% which in our notation corresponds to
% $\Phi=F_1$ and $M=2F_4$.  The decomposition of the curvature
% term further gives
% $F_2=-F_1/4$, $F_3=0$, and $F_5=2F_{1,\phi}$.
% Our curvature descent equation,
% \begin{equation}
% U_\phi F_{1,\phi}-\frac14UF_1+1=0,
% \end{equation}
% is therefore the three-dimensional version of the usual dBVV \cite{deBoer:1999tgo}
% second-order Hamilton-Jacobi relation, while the accompanying relation
% between $F_4$ and $F_{1,\phi}$ reproduces the corresponding scalar-gradient
% constraint.  The finite-temperature system
% \eqref{eq:F1thermalSimple}-\eqref{eq:F5thermalSimple} can thus be viewed
% as a hydrostatic extension of the dBVV descent equations: thermal
% anisotropy splits the two covariant zero-temperature structures into five
% stationary structures, which recombine into the standard covariant form as
% $\eta\rightarrow0$.

% ------------------------------------------------------------
\subsubsection*{Check of the remaining Einstein-scalar equations}
% ------------------------------------------------------------

In radial Gaussian gauge, the Hamiltonian and momentum constraints are
supplemented by
\begin{align}
\dot K_{\mu\nu}
={}&
R_{\mu\nu}[\gamma]
+
2K_{\mu\rho}K^\rho{}_\nu
-
KK_{\mu\nu}
-
\frac12\nabla_\mu\phi\nabla_\nu\phi
-
\frac12V\gamma_{\mu\nu},
\\
0={}&
\ddot\phi+K\dot\phi+D^2\phi-V_\phi .
\end{align}
Substituting the characteristic flow generated by the Hamilton-Jacobi
functional into these equations gives no additional independent condition.

Thus the Hamiltonian constraint fixes the local coefficient functions. 
The momentum constraint selects the hydrostatic branch, and the remaining
Einstein-scalar equations provide a consistency check of the resulting
radial evolution.

\section{ADM decomposition for time-like evolution}
\label{time evolution appendix}
In the previous subsection, we discussed the radial evolution. Now, we are going to discuss the time evolution. This is relevant for discussing the interior of the black hole, where the radial direction is time-like. The metric ansatz is
\begin{eqnarray}
    g_{\mu\nu}dx^{\mu}dx^{\nu}=-N^2 dt^2+\gamma_{ij}(dx^i+\beta^i dt)(dx^j+\beta^j dt)).
\end{eqnarray}
We start with the Einstein-Hilbert action
\begin{eqnarray}
    S= \int d^4x \, \sqrt{-g}\, {}^4 (R- 2 \Lambda).
\end{eqnarray}
Here, we take the following normal vector
\begin{eqnarray}
    n^{\mu}=(\frac{1}{N}, -\frac{\beta^i}{N}).
\end{eqnarray}
One can also define the normal evolution vector labeled as $m$
\begin{eqnarray}
    m:= N \, n, \quad m.m=-N^2
\end{eqnarray}
Then the Lie derivative of the induced metric can be written as
\begin{eqnarray}
    \mathcal{L}_{m} \gamma= -2 N\, K \implies K= -\frac{1}{2} \mathcal{L}_n \gamma.
\end{eqnarray}
With the above ADM decomposition of the metric, we can write the 4d Ricci scalar in terms of the 3d Ricci scalar as follows
\begin{eqnarray}
    {}^4 R= R+K^2+K_{ij} K^{ij}-\frac{2}{N} \mathcal{L}_m K-\frac{2}{N} D_i D^i N
\end{eqnarray}
The determinant can be written as $\sqrt{-g}= \sqrt{\gamma} N$, and then the action can be written as
\begin{eqnarray}
    S=\int \sqrt{\gamma} d^4 x\, \Big[N(R+K^2+K_{ij}K^{ij})-2 \mathcal{L}_m K- 2 D_iD^i N\Big]
\end{eqnarray}
After some manipulations and carefully treating boundary terms, we yield \cite{Gourgoulhon:2007ue} \footnote{ We have dropped all the total derivative terms (see section 4.5.1 of \cite{Gourgoulhon:2007ue} for more details). }
\begin{eqnarray}
    S=\int dz \int \sqrt{\gamma}\,  d^3 x\,\, \Big[N(R-K^2+K_{ij}K^{ij}-2 \Lambda)\Big]
\end{eqnarray}

\textbf{Hamiltonian approach:}\\

The above action is a functional of configuration variables $q=(N,\beta^i,\gamma_{ij})$ and its  derivative $q'=(\gamma_{ij}',N',\beta^{i'})$. Then the Lagrangian density for the above action in terms of configuration variables $q=(\gamma_{ij},N,\beta^i)$ can be written as 
\begin{eqnarray}
    L(q,\dot{q})= N \sqrt{\gamma} (R+K_{ij}K^{ij}-K^2- 2 \Lambda)= N \sqrt{\gamma}  \Big[R-2 \Lambda+(\gamma^{ik}\gamma^{jl}-\gamma^{ij}\gamma^{kl}) K_{ij} K_{kl}\Big]\nonumber\\
\end{eqnarray}
The extrinsic curvature $K_{ij}$ is
\begin{eqnarray}
\label{kijmet}
    K_{ij}=\frac{1}{2 N} (\gamma_{ik} D_j \beta^k+\gamma_{jk} D_i \beta^k- \dot{\gamma}_{ij}).
\end{eqnarray}
The Lagrangian doesn't depend on the $\dot{N}$ and $\dot{\beta}^i$. Hence, these are not dynamic variables. The induced metric $\gamma_{ij}$ is the only dynamical variable.
The canonical momenta conjugate to the $\gamma_{ij}$ is
\begin{eqnarray}
\label{pijmet}
    \pi^{ij}:= \frac{\partial L}{\partial \dot{\gamma}_{ij}}=(K \gamma^{ij}- K^{ij})\sqrt{\gamma}
\end{eqnarray}
%From the above Lagrangian and writing $K_{ij}$ in-terms of the induced metric \eqref{kijmet}, we get 
%\begin{eqnarray}
 %   \pi^{ij}= \sqrt{\gamma}(K \gamma^{ij}- K^{ij})
%\end{eqnarray}
The other conjugate momenta to $N$ and $\beta^i$ are zero since they are not dynamical.
\begin{eqnarray}
    \pi^N:= \frac{\partial L}{\partial \dot{N}}=0, \quad \pi_i^{\beta}:= \frac{\partial L}{\partial \dot{\beta}^i}=0
\end{eqnarray}
The Legendre transform of the Lagrangian gives the Hamiltonian density \footnote{Here again we have thrown away total derivative terms like $D_j(K \beta^j - K_i^j \beta^i)$. } 
\begin{eqnarray}
    \mathcal{H}&&= \pi^{ij} \dot{\gamma_{ij}}-L\nonumber\\
  &&=-\sqrt{\gamma}\Big[N\, C_0- 2 \beta^i C_i\Big] 
\end{eqnarray}
Here 
\begin{eqnarray}
    &&C_0:=R +K^2- K_{ij}K^{ij}+ 2 \Lambda\\
    &&C_i:=D_j K^{j}_i- D_i K
\end{eqnarray}
The total Hamiltonian now becomes
\begin{eqnarray}
    H= \int_{\Sigma} \, \mathcal{H} \, \,  d^3 x
\end{eqnarray}

The Extrinsic curvature $K_{ij}$ can be written in terms of conjugate momenta by inverting the relation \eqref{pijmet} 
\begin{eqnarray}
    K_{ij}= K_{ij}[\gamma,\pi]= \frac{1}{\sqrt{\gamma}}(\frac{1}{2} \pi \gamma_{ij}-  \pi_{ij})
\end{eqnarray}
Here, the induced metric $\gamma_{ij}$ is used for raising and lowering the indices, and then the trace of momenta is written as $\gamma_{kl} \pi^{kl}=\pi$.
 The Hamilton equation of motion can be found by variation. The first equation can be written as 
\begin{eqnarray}
   \frac{\delta H}{\delta\, \pi^{ij}}= -2 N K_{ij}+D_i \beta_j+D_j \beta_i= \dot{\gamma}_{ij}.
\end{eqnarray}
Other equations are
\begin{eqnarray}
    \frac{\delta H}{\delta \gamma_{ij}}= - \pi^{ij\,'}, \quad \frac{\delta H}{\delta N}=-C_0=0, \quad \frac{\delta H}{\delta \beta^i}= 2 C_i=0
\end{eqnarray}
Subsequently, one can add matter.

    \section{Ward identity for Broken conformal symmetry}
    \label{ward identity}
In this section, we derive the broken Ward identity due to placing the theory on a non-trivial background, i.e., on a thermal manifold $S^1_{\beta} \times R^{d-1}$. Some of the conformal symmetries will be broken due to the finite temperature $T$, which is a scale. The non-triviality of the manifold leads to the modification of the standard flat space Ward identity. Here, we explicitly derive the Ward identity for the broken dilatation symmetry. But it can be generalized for the other generators, like Lorentz boosts and special conformal transformations. The goal of HJ theory is to reproduce these results.  \\

Here we will review the work of Marchetto et al. \cite{Marchetto:2023fcw}.  We start with theory on $S^1_{\beta} \times R^{d-1}$. The infinitesimal symmetry transformation represented as $\phi'(x)= \phi(x)- i \omega_a G_a \phi(x)$. Here, the variation can be written as
    \begin{eqnarray}
        i \omega_a G_a \phi(x)= \omega_a \frac{\delta x^{\mu}}{\delta \omega_a} \partial_{\mu} \phi- \omega_a \frac{\delta F}{\delta \omega_a}
    \end{eqnarray}
    Here $F$ represents the variation of the field $\phi$ under the symmetry generator $G_a$. Under the conformal transformation, the correlation function remains invariant.
    \begin{eqnarray}
    \label{Tcorr}
     \langle O_1(x_1)\cdots O_n(x_n)\rangle_{\mathcal{M}}=\frac{1}{Z}\int [d \phi']\Big( O_1(x_1)+\delta O_1(x_1)\cdots O_n(x_n)+\delta O_n(x_n)\Big) e^{-S'[\phi']}  \nonumber\\ 
    \end{eqnarray}
  The change in the action is
    \begin{eqnarray}
        S'[\phi']=S[\phi]+\int_{\mathcal{M}} d^d x \sqrt{g}\, \nabla_{\mu} \omega_a(x) J^{\mu}_a(x).
        \end{eqnarray}
        The currents $J_{a}^{\mu}$ in a Lagrangian theory can be found as
        \begin{eqnarray}
            J^{\mu}_a&&=\frac{\delta x^{\mu}}{\delta \omega_a} \mathcal{L}- \frac{\partial \mathcal{L}}{\partial \partial_{\mu}\phi}\partial_{\nu} \phi \frac{\delta x^{\nu}}{\delta \omega_a}+\sum_{\phi}\frac{\partial \mathcal{L}}{\partial \partial_{\mu}\phi} \frac{\delta F}{\delta \omega_a}\nonumber\\
            &&=T^{\mu\nu}_{can}\frac{\delta x_{\nu}}{\delta \omega_a} +\sum_{\phi}\frac{\partial \mathcal{L}}{\partial \partial_{\mu}\phi} \frac{\delta F}{\delta \omega_a}.
        \end{eqnarray}
        Now we can insert the transformation of fields and action into the correlation function \eqref{Tcorr}. After expanding it to the first order in $\omega_a$, we get
    \begin{eqnarray}
    \label{wardbroken}
        \sum_i \langle O_1(x_1)\cdots \delta O_i(x_i)\cdots O_n(x_n)\rangle_{M}&&= \langle \delta S\, O_1(x_1)\cdots O_n(x_n)\rangle_M\\
        &&= \int_M d^d x \sqrt{g} \nabla_{\mu} \omega_a(x) \langle J^{\mu}_a(x) O_1(x_1) \cdots O_n(x_n)\rangle_M\nonumber\\
    \end{eqnarray}
    Here $\delta O(x)= -i \omega_a (x) G_a O(x)$. We can do the functional derivative on both sides with respect to $\omega_b(y)$. 
    \begin{eqnarray}
        -i \langle \delta(x_i-y) \langle O_1(x_1)\cdots G_a O_i(x_i)\cdots O_n(x_n)\rangle_{M}=\int_M d^d x \sqrt{g}\, \partial_{\mu} \delta^{d}(x-y)\langle J_a^{\mu}(x) O_1(x_1)\cdots O_n(x_n)\rangle_M\nonumber\\
    \end{eqnarray}
    Then the un-integrated Ward identity can be written as
    \begin{eqnarray}
      i \langle \delta(x_i-y) \langle O_1(x_1)\cdots G_a O_i(x_i)\cdots O_n(x_n)\rangle_{M}=  \nabla_{\mu}^y \langle J_a^{\mu}(y) O_1(x_1)\cdots O_n(x_n).\rangle_M \nonumber\\   
    \end{eqnarray}
      As an example, for a one-point function,
    \begin{eqnarray}
        i \delta(x-y)\langle[D,O](x)\rangle_{\beta}= \frac{\partial}{\partial y^{\mu}}[(y-x)_{\nu}\langle T^{\mu\nu}(y)O(x)\rangle_{\beta}]\nonumber\\
    \end{eqnarray}
    We can integrate both sides of the equation and get
    \begin{eqnarray}
      i \sum_i \langle O_1(x_1)\cdots G_a O_i(x_i)\cdots O_n(x_n)\rangle_{\beta}= \int d^{d-1}y \int_0^{\beta} dy_0 \frac{\partial}{\partial y^{\mu}} \langle J_a^{\mu}(y)O_1(x_1)\cdots O_n(x_n)\rangle_{\beta}\nonumber  \\
    \end{eqnarray}
    The right-hand side is zero if the theory is defined on $R^d$. This is the standard Ward identity for conformal invariance.
    \begin{eqnarray}
        i \sum_i \langle O_1(x_1)\cdots G_a O_i(x_i)\cdots O_n(x_n)\rangle_{R^d}=0
    \end{eqnarray}
    but the right-hand side doesn't vanish when integrated over $S^1_{\beta}\times R^{d-1}$. We can evaluate the breaking term in terms of 
    \begin{eqnarray}
    \label{broken}
        \int_{R^{d-1}} d^{d-1}y \langle[J_a^{0}(\beta,y)-J_a^0(0,y)]O_1(x_1)\cdots O_n(x_n)\rangle_{\beta}&&\equiv  \int_{R^{d-1}} d^{d-1}y \langle\Gamma_a^{\beta}(y) O_1(x_1)\cdots O_n(x_n)\rangle_{\beta}\nonumber\\
        &&=\beta \int_{R^{d-1}} d^{d-1}y \langle T^{00}(y) O_1(x_1)\cdots O_n(x_n)\rangle_{\beta}\nonumber\\
    \end{eqnarray}
    here $\Gamma_a^{\beta}(y)=J_a^{0}(\beta,y)-J_a^0(0,y) $. For the Dilatation operator this becomes $\Gamma_a^{\beta}(y)=\beta T^{00}$. At finite temperature, the correlation function is invariant under these sets of scaling transformations as
    \begin{eqnarray}
        O_i \rightarrow \lambda^{-\Delta_i}O_i, \quad x_i \rightarrow \lambda_i x_i,\quad \beta \rightarrow \lambda_i \beta
    \end{eqnarray}
    Intuitively, the scaling of the temperature is equivalent to the scaling of the metric $g_{00}$. The stress tensor component $T^{00}$ is the source of that scaling. Hence, for one one-point function, we have
    \begin{eqnarray}
        \frac{\partial}{\partial \beta} \langle O\rangle_{\beta}=-\frac{1}{\beta}\int d^{d-1}x \int_0^{\beta} d\tau \langle T^{00}(\tau,\vec{x}) O(0)\rangle_{\beta}= - \int d^{d-1}x \langle T^{00}(\tau,\vec{x}) O(0)\rangle_{\beta}\nonumber\\
    \end{eqnarray}
    Hence, for the broken Ward identity terms \eqref{broken} can be written in terms of the derivative with respect to $\beta$. It becomes
    \begin{align}
    \label{boxedeq}
    \boxed{
        (D+\beta \frac{\partial}{\partial \beta})\langle O_1(x_1)\cdots O_n(x_n)\rangle_{\beta}=0
        }
    \end{align}
    The breaking of the dilatation Ward identity is compensated by variation with respect to the $\beta$. The goal of the HJ theory is to reproduce this Ward identity.\\

\section{WdW equation and Weyl anomaly}
\label{app:new}
In this appendix, we write the WDW equation \footnote{A word on terminology:- Wheeler de-Witt equation is a functional equation acting on wavefunction for quantum gravity, while Hamiltonian constraints are purely classical constraints often written in metric variables. These canonical variables become operators in quantum mechanics, and then the wavefunctions should respect the classical constraints.} without matter in AdS$_3$. 
\begin{eqnarray}
    \Big(K^{ij} K_{ij}-K^2- R [\gamma]+2 \Lambda\Big) \Psi[\gamma]=0.
\end{eqnarray}
The conjugate momenta can be written in terms of extrinsic curvature. Then the constraint equation becomes
\begin{eqnarray}
    \Big(\frac{(16 \pi G)^2}{ \mathrm{det} \gamma} (\Pi^{ij} \Pi_{ij}- \Pi^2- R - 2 \Lambda)\Big) \Psi[\gamma]=0
\end{eqnarray}
Furthermore, one can define a new wavefunction as 
\begin{eqnarray}
    \Psi= \exp(\frac{1}{8 \pi G \ell} \int d^2 x \sqrt{\gamma}) \hat{\Psi}
\end{eqnarray}
This changes the conjugate momenta as 
\begin{eqnarray}
    \Pi^{ij} \rightarrow \Pi^{ij}- \frac{i }{16 \pi G \ell} \sqrt{\gamma} \gamma^{ij}.
\end{eqnarray}
The transformed wavefunction now satisfies 
\begin{eqnarray}
    \Bigg(\frac{i}{\sqrt{\mathrm{det} \gamma}} \Pi +\frac{8 \pi G \ell}{det \gamma} (\Pi^{ij}\Pi_{ij}-\Pi^2)-\frac{\ell}{32 \pi G} R\Bigg) \hat{\Psi}=0
\end{eqnarray}
The second term is irrelevant at large distances. Hence, at leading order at large distance, the WDW equation becomes
\begin{eqnarray}
    \Bigg(\frac{i}{\sqrt{\mathrm{det} \gamma}} \Pi -\frac{\ell}{32 \pi G} R\Bigg) \hat{\Psi}=0
\end{eqnarray}
The trace of conjugate momenta $ i \, \Pi$ is the generator of the Weyl transformation of the metric because by definition $\pi^{ij}=-i\frac{\delta}{\delta h_{ij}} $. Hence, the above WdW equation can be identified as the anamalous Ward identity of 2d CFT.

\bibliographystyle{JHEP}
\bibliography{ref}
\end{document}